\documentclass[%
 reprint,
 amsmath,amssymb,
 aps,
]{revtex4-2}

\usepackage{graphicx}% Include figure files
\usepackage{dcolumn}% Align table columns on decimal point
\usepackage{bm}

\usepackage{amsmath}

\DeclareMathOperator*{\argmax}{\arg\!\max}

\usepackage{float}

\usepackage{hyperref}

\usepackage[nobiblatex]{xurl}

\begin{document}

\preprint{APS/123-QED}

\title{Improving Punching ``Power":\\Reconsidering the roles of initial strike velocity and effective mass}

\author{Joseph L. Natale}
\affiliation{
 Fight Science Research Division, Techanique Consulting, LLC
}

\date{June 11, 2026}

\begin{abstract}
We review the principal features of one-dimensional, elastic collisions pertinent to modeling strike-mediated impacts in martial arts and combat sports contexts. Arguing for the post-collision velocity of a stationary, target mass as a useful proxy for determining overall punch efficacy, we demonstrate that pre-collision velocities for real strikes are bounded, such that \emph{speed prior to impact} has a more limited effect on the impulse experienced by said target than suggested by supporting literature.
We find that the value of incoming, ``effective" mass -- despite eluding uniform establishment criteria or measurement protocols across the same body of literature -- exerts an independent influence on the target velocity which can, in principle, match or exceed that of the strike speed. We perform explicit numerical analyses to supplement these theoretical considerations, in dynamical regimes relevant to real 
human combat, and then briefly discuss the potential for a given fighter to attain a fold-increase in effective mass (from a reasonable baseline value based on bodyweight, and supported by previous work) sufficient to realize limits for upper-limb strikes to the body, in practice, as a trainable goal.
\end{abstract}

\maketitle

%\tableofcontents

\section{\label{intro}Introduction}

In both recreational martial arts and modern combat sports, across cultures and stylistic variants, the practical potencies of individual striking techniques are given a considerable level of attention -- despite storied histories of disagreement over the principal sources of ``punching power"~\cite{bluestein2014research}. In traditional, especially Eastern, circles, the blending of dialectic discourse and individual discovery, often done in %the 
explicit absence of quantitative metrics and contemporary scientific terminology~\cite{bluestein2014research,lee1975tao}, is an expected part of learning specific ``forms." This can lead to a more diffuse, interpreted understanding of what renders strikes effective during `real' fighting~\cite{quinn1990bouncer,quinn1996real}. On the other side of this coin, contemporary promotion companies hosting professional fight competitions, like the Ultimate Fighting Championship (UFC), place a more unambiguous and disproportional emphasis on effective, powerful punching (via score-valuation of ``significant" strikes~\cite{mmajudging2016} and financial incentives on wins by knockout~\cite{eisinger2018,gift2019performance,whitebonus2026}), gathering in excess of one hundred sixty individual metrics, per bout, to drill down on win statistics and trainable qualities~\cite{ufcpi2018}.

Despite new levels of measurement precision offered by modern kinesiology, sports medicine, and fight science, a clear, operational definition for common colloquial terms, such as ``impact" of a strike, continue to elude studies on contributors to fight-relevant punching performance~\cite{lenetsky2013assessment,monfared2021contributing, beattie2022role, pinto2025influence}. An inherent, implicit difficulty associated with these assessments stems from the fact that individual studies vary tremendously in their outcome metrics and intended application venues. For instance, whereas one investigation might be concerned with fighter safety and the prevention of concussions (i.e., minimization of acceleration when striking a hard, relatively target like the skull), another might involve the peak forces incurred by a more deformable target, coupled to a floor, to gain insight into strikes thrown within ground-control, grappling contexts (i.e., ``ground \& pound" techniques)~\cite{khatib2024brain,beranek2020upper,beranek2022performance}.

Regardless of the particular application, the two most salient mechanical variables with robust, tractable, physical mappings to a broad range of striking outcomes are the \emph{velocity} and \emph{effective mass} of the striking implement (e.g., fist or foot), with the latter having experienced an increased prominence in the literature of the past decade while also submitting to far more inhomogeneous standards and measurement methodologies than the former. Although the concept of an ``effectively participating," less-than-full fraction of the striker's mass being available to interact with a given target remains interpretable in principle -- a notion that may have been approximated by karate and boxing exponents for decades, even centuries, by other names (consider, for instance, the analogous but less precise scope of ``kime"~\cite{mcgill2012kettlebell} and the pejorative term ``arm punch," versus ``full-body strike," in conversations about how to maximize mass~\cite{lee2017effect}) -- the widespread adoption of gold-standard protocols and criteria to define a unique variable ``$m_{eff}$", including standardized measurement apparatus designs, is not yet a realized feature of mainstream research or `best-practices' in striking~\cite{lenetsky2015effective}.

In addition to these disparities -- which otherwise might be excused, given the sheer time and unanimity required for developing a novel technical definition -- one available review, and a meta-study~\cite{lenetsky2015effective,khatib2024brain}, reveal that a majority of ``effective mass" estimations (in a stark, ironic contrast to admonitions against utilizing so-called ``arm strikes") report values falling within $5$-$10\%$ of fighter body masses, equivalent to segmental masses for single, upper limbs. If effective mass values for professional fighters are bounded to this order of magnitude, the sensitivity of striking outcomes to $m_{eff}$ as a variable -- and the relevance of $m_{eff}$ as an incrementable, trainable quantity -- vanishes.

In general, values for effective mass have been assessed in several ways~\cite{lenetsky2015effective}. The most basic of these is a solution of an inverse-mechanics problem, whereby masses, final velocities, and incoming strike velocities are measured directly for stationary targets, such that incoming, effective masses can be found via conservation laws. Since targets differ across studies in their material compositions (and therefore stiffness and compliance) and support structure (e.g., a \emph{fixed} target, mounted on a bracket or wall several orders of magnitude larger than the incoming $m_{eff}$, versus a \emph{movable}, hanging heavy bag), one cannot expect to rate strong concordances among approaches to mapping the ``objective" data they produce to the practical advice those studies inspire~\cite{beranek2023force,uthoff2023review}. Opinions generated across these research articles also may not mesh with the subjective opinions of striking coaches, which are already diverse and nonuniform despite their track records and subject matter expertise. Still, anecdotal evidence suggests that coaches can ``see" loose proxies for effective mass in powerfully executed strikes, without calculations~\cite{mcgill2010evidence}.

To illustrate, a trainer might pay attention to whether a formerly-stationary heavy bag swings, versus ``jumps," as a consequence of strike-contact; whether they would ascribe an ephemeral quality of ``snap" to the collision interaction; or, what happens to the body of a partner who holds a lower-mass target, such as a pad or mitt. Qualitative observations by veteran coaches can lack modern scientific precision and may even combine or conflate multiple physical quantities, but also boast \emph{face validity} -- and success of trainees as a ``validation" -- nonetheless. 

Beyond the aforementioned difficulties associated with relying on \emph{effective mass} as an explanatory variable for perceived power of a strike -- for, indeed, the only opinion-authority of import should be the receiver of the strike! -- certain camps and schools of thought maintain that the \emph{velocity} of the incoming implement dominates any mass-effect in determining ``power," at least in the colloquial sense. Putting aside the broader difficulties in reconciling the wider, popular or na\"{i}ve repertoire of pseudoscientific terminology with standard mechanical nomenclature -- a task beyond the scope of this article -- there appear a handful of routinely referenced `physical' facts that confound such discussions. A token example~\cite{ishac2021evaluating} is that the kinetic energy of a body, while directly proportional to the body's mass, depends on the \emph{square} of the magnitude of its velocity; the common argument, which we will examine below, is that the \emph{speed} of the implement retains a greater ultimate importance -- a greater predictive value -- than mass, in 
development of practically-effective strikes. 

Here, we re-examine several fundamental aspects of physically-modeled strike collisions that underlie diverse, various possible operational definitions of effective mass that have been explored in recent literature -- with a critical eye. We begin by reviewing the simplest conceivable case of a single moving point mass, set to strike a stationary target of similar or disparate mass, and examine the outcomes from the dynamical systems perspective, in the regimes that applies to real, human-sized combatants. 

We articulate that the contribution of the velocity term to strike outcomes can, indeed, be dominated by that of the mass term, against what is popularly assumed, and argue in favor of at least one clear, qualitative indicator as a practically-deployable proxy for the ratio of the effective mass of the striker to that of the target. In response to the above methodological issues stemming from the difficulties associated with measuring effective mass, we keep comments brief, and then direct readers to~\cite{natale2026a}. We conclude with an acknowledgment of physical limitations on the amount of effective mass (relative to striker bodyweight) that can be imparted to, or reasonably interact with, a stationary target; we postpone a solution -- a saturation of these limits
-- to ongoing, future work~\cite{natale2026_shukjpotentialM}.

\section{Dynamical Analysis for a Stationary Target}
\label{setup}

\subsection{One-Dimensional, Elastic Collision}
\label{elastic_equations}
Consider a point mass $m$, subject to no external forces, traveling with an initial velocity $v$ on a straight-line path. If this mass comes into contact with a second point mass $M$, of arbitrary size, initially at rest -- and we can assume the ensuing collision is perfectly elastic -- both undergo changes in motion, which can be ascertained by applying conservation of momentum and conservation principles.

From the conservation of momentum, we have

\begin{eqnarray}
mv ~= ~mv_f + MV_f
\label{eq:1},
\end{eqnarray}

where $v_f$ describes the velocity of the initially-moving mass $m$ immediately following the impact and $V_f$ denotes the velocity taken on contemporaneously by the initially-stationary mass $M$. Similarly, by conservation of (kinetic) energy, we can write:

\begin{eqnarray}
mv^2 ~= ~m{v_f}^2 + M{V_f}^2
\label{eq:2},
\end{eqnarray}

where all constant factors of $\frac{1}{2}$ have been omitted for brevity and clarity. This pair, or ``system," of equations has a well-known solution, wherein the joint dependencies on $m$ and $M$ emphatically \emph{separate} from the effects of $v$:

\begin{eqnarray}
v_f ~=~ \left(\frac{m-M}{m+M}\right)v\;,
\label{eq:3}
\\
V_f ~=~ \left(2\frac{m}{m+M}\right)v%
\label{eq:4}.
\end{eqnarray}

Introductory physics textbooks~\cite{wolfson2020essential} often
segment this solution into several, separate ``edge cases" describing the motion of the two masses. If \emph{i)} the initially-moving mass $m$ is so much smaller than the at-rest mass that its value can all but be neglected ($m\ll M$), this smaller mass will reverse its direction of travel but keep nearly all its initial kinetic energy: its post-collision speed will be nearly equal to $v$, but in the opposite direction. The large mass $M$ will remain essentially unperturbed, accruing minimal momentum, along the vector of impact. Vice versa, when \emph{ii)} the initially-moving mass dominates its stationary target -- that is, in spite of what is now a somewhat deceptive nomenclature, $m\gg M$ -- it continues at almost speed $v$, virtually unimpeded by its would-be obstacle $M$, which in turn accelerates to a velocity approaching $+2v$.

When both masses are comparable \emph{iii)} -- indeed, when they are identical in value -- the incoming mass $m$ ceases moving entirely, in the frame of reference used to analyze the problem here; the target picks up its full store of momentum and kinetic energy. This underlies the familiar phenomenon of the ``stop shot" in the game of billiards, in which a cue ball (then called \emph{dead ball}) desists in moving any farther upon impact, and transfers all its quantity of
motion to a nearly equal-mass recipient of impact.

We must distinguish between the three key
situations, in terms of their relevance to striking: $m<M$, $m>M$, and $m\approx M$. We interpret them as follows.

Let $m$ represent the \emph{effective mass} associated with the incoming strike itself (regardless of the measurement paradigm; clearly, effective mass does not exist as a static quantity on its own, but only has meaning as an explanatory or constitutive variable in process of a collision) and $M$ that of the target (analogously, this may not be equal to the full, statically-measured target mass because the fraction of total mass that participates in the collision depends on how strongly the mass components are coupled to 
each other, and even to their supporting surfaces).

\begin{itemize}

    \item When $m<M$, and emphatically \emph{not} exclusively when $m\ll M$, the incoming mass $m$ -- which can be thought of as representing the striking implement that actually makes contact, plus a further fraction of the striker's body -- will recoil upon impact. This is an inescapable consequence of Equations~\ref{eq:3}-\ref{eq:4}.
    
    \item If instead $m>M$ -- again, a weaker condition than $m\gg M$ -- the striker's incoming mass will continue, albeit abated, with a velocity smaller in magnitude than $v$. Ignoring its precise magnitude, the target's post-collision velocity $V_f$ will exceed $v$, the original velocity of the incoming strike, unlike the previous case above (whereupon $m$ induces a positive target velocity by making the sacrifice of recoil as a compromise, or payoff; this affects impact forces~\cite{natale2force}).
    
    \item The (rarer) scenario in which $m=M$ exactly is the \emph{only} case in which implicated, striking mass $m$ will come to a full stop, with no recoil and zero residual motion. If there were truly \emph{zero} detectable reversal, or continuation of motion, we could be certain that 
    the effective masses, both striker and target, match.
    
\end{itemize} 

Recapitulating, although the initial (pre-collision) velocity of the mass $m$ contributes to the \emph{magnitude} of both ultimate velocity values, $v_f$ and $V_f$, it has no effect on the \emph{qualitative} outcome, in terms of the \emph{relative signs}, or vector directions, attributed to either. Since the $v$ term factors out of both Equations~(\ref{eq:3}-\ref{eq:4}), it only affects their individual values quantitatively, but does not affect their \emph{ratio} at all. We assert that this observation has practical implications for assessing striking outcomes in fighters.

\subsection{Increasing Initial Velocity and Striking Mass}
\label{effects_m_v}

The above contentions do not imply that adjusting the initial velocity $v$ of the striking mass yields a vanishingly small effect on $v_f$ and $V_f$. Rather, it is evident that, e.g., doubling the value of this initial velocity $v$ will double the values of both $v_f$ and $V_f$, whereas it can be calculated that doubling the effective striking mass $m$, even relative to the special case $m=M$, leads much more modestly to $v_f = 0.3v$ instead of $v_f=0$ and $V_f\rightarrow1.3V_f$, respectively.

This particular evaluation, $m\rightarrow 2M$, might serve as a valid description of the difference in outcomes if a fighter whose effective mass \emph{matches} that of the target \emph{swapped places} with another, in a heavier weight class, somehow capable of mustering twice the effective mass in collision -- or, if the fighter suddenly acquired a way to incorporate twice their previous mass-fraction. 
Taking the ratio 

\begin{eqnarray}
\frac{2\frac{(2m)}{(2m)+M}v}{2\frac{m}{m+M}v}
\label{eq:5},
\end{eqnarray}

describing the fold-change in $V_f$ for the target mass $M$ when \emph{arbitrary} incoming mass $m$ is doubled, leads to

\begin{eqnarray}
\frac{V_f,~new}{V_f,~old}~=~ 2\cdot\frac{m+M}{2m+M} ~=~ 2 - \frac{2m}{2m+M}
\label{eq:6},
\end{eqnarray}

or that $V_f\rightarrow2V_f$ (minus first-order correction $\mathcal{O}( \frac{-2}{~M} )$) when $m\rightarrow2m$, under the initial condition $m\ll M$. On the other hand, if $m\gg M$, doubling the mass from $m\rightarrow 2m$ leaves the post-collision target velocity $V_f$ materially unchanged, as it should be expected. Intermediate values of the ratio $u\equiv\frac{m}{M}$ boost $V_f$ by \emph{less} than two-fold.

In general, for a $c$-fold increase in the measured value of the incoming effective strike mass (that is, a replacement $m\rightarrow cm$ with an otherwise equivalent, but larger, mass), the corresponding fold-increase in $V_f$ is given by  

\begin{eqnarray}
k~\equiv~\frac{cm+cM}{cm+M}~~=~~\frac{c\left(u+1\right)}{cu+1}
\label{eq:7}.
\end{eqnarray}

We can tease apart the dependence of $k$ on $c$, versus its dependence on $u$, via factorization and cancellation:

\begin{eqnarray}
k~=~\frac{\left(u+1\right)}{u+\frac{1}{c}}
\label{eq:7b}.
\end{eqnarray}

Numerical evaluation  of Eq.~(\ref{eq:7b}) confirms that, for small $u$, the response in $V_f$ that results from making the change $m\rightarrow cm$ remains linear: $V_f\rightarrow cV_f$, as for the special case of $c=2$ discussed above. However, such linear increases still cannot compete with the directly-proportional, fold-increases in $V_f$ that can be achieved, in principle, by scaling the incoming strike speed $v$ by an arbitrary constant factor $b$ -- as an independent effect -- for several reasons.

Why is it so much more trivial to increase the value of $V_f$ via velocity tuning ($v\rightarrow bv$), as opposed to adjusting the incoming, effective mass $m$? Note that the velocity term $v$ factors out of the expression for $V_f$ entirely -- it has no parametric dependence on $m$ or $M$ -- and the linearity of its effect on $V_f$ persists across its all accessible values in its domain. The mass transformation ($m\rightarrow cm$) induces a direct-proportionality effect on $V_f$ only when $m$ is \emph{small} in comparison with $M$; with such minuscule values of the incoming effective mass that $u\approx 0$, multiplying the value of $V_f$ by a constant factor $c$ still leads to $V_f\approx 0$, via Eq.~\ref{eq:4}.

Nevertheless, the validity of all these remarks hinges on the extent to which offensive strikes and opposing, human combatants can be modeled, with accuracy, as elastically-interacting, ``point" masses. For the moment, we suspend all questions on the validity of our assumptions regarding elasticities and point particles to focus more precisely on the \emph{dynamical regimes} over which the quantities $v$ and $u$ -- and, thus, $V_f$ -- can reasonably be expected to vary, in the contexts of realistic combat. This change of focus is to highlight the second reason why $v$ \emph{appears} to have greater influence on $V_f$: we have so far avoided situations in which $m<M$, but $u\gg 0$. Only in this ``intermediate" regime can $k$ take values to rival the velocity-multiplier $b$, as an independent effect. This will become our focus.

\subsection{Insufficiency, or Non-Domination, of 
strike velocity in determining post-collision, target speed}
\label{insufficiency_v}

To summarize the key considerations from the previous section, a multiplicative adjustment to the initial \emph{velocity} of the incoming mass $m$, such that $v\rightarrow bv$, where $b$ is some constant, leads us to $V_f\rightarrow bV_f$. Meanwhile, scaling the value of the incoming effective mass $m$ by $c\neq b$ exerts a more subtle, potentially smaller-magnitude, effect.

Typical strike velocities seen in professional boxing and mixed martial arts (MMA) settings~\cite{kimm2015hand,corcoran2024impact,khatib2024brain} are on the order of $v\sim 8-10~\frac{m}{s}$. 
According to the Unified Rules of Mixed Martial Arts~\cite{unifiedrules2024}, male UFC fighters tend to vary between an appreciable minimum of 125, and a maximum of 260, pounds -- approximately a 2-fold range -- with the median weight class across all fights in a typical schedule lying close to the allowable upper limit of the Lightweight class ($155$ pounds $\approx70kg$)~\cite{nichols2024project,fightmatrix2026records}. Two-fold variation in the mass of a single, upper limb~\cite{plagenhoef1983anatomical,zatsiorsky1983mass}, as an ansatz for the effective mass of a strike, suggests $m\sim[3.5,7]~kg$.

Based on Eq.~(\ref{eq:7b}), these statistics may seem insufficient to cause shifts in the achievable target velocity $V_f$ as large as those that can be realized by augmenting the velocity $v$. This logic will break down quickly once we consider \emph{i)} the percentage of body mass that would typically participate in modeled collisions, using estimates from previous work, and \emph{ii)} the ranges over which strike velocities tend to vary, considering amateur fighters and professionals.

For the interacting-mass proportion, we note that studies across boxing, kung fu, karate, and MMA -- each style with idiosyncratically different techniques and execution patterns -- routinely arrive at estimates of $m\sim 5kg$, and sometimes less, for those effective mass values associated with their analyzed strikes~\cite{lenetsky2015effective,khatib2024brain}. Even the highest end of these figures, reported in only a single paper series~\cite{kacprzak2025biomechanics} (based on searches near the time of writing), reach only $m\sim30kg$ -- barely reaching 50\% of the lowest `regulated' body mass -- equivalent to an average torso mass~\cite{plagenhoef1983anatomical,zatsiorsky1983mass} for fighters in most ``male" divisions (NB: \emph{female} fighters in the UFC are restricted to 145 pounds, or $<66 kg$).

Hence, in accordance with the training procedures and techniques that dominate the modern combat scene, the incoming striking mass can indeed, at least in theoretical terms, \emph{double}, \emph{triple}, or even \emph{quadruple} from its baseline value $m\sim[5,10]~kg$ -- i.e., $c\in[2,4]$ -- if a fighter is able to incorporate \emph{more} than just a single limb into the strike -- and if the aforementioned, reported value of $\sim30kg$ can be taken as an accurate representation of an upper limit. Still, even several-fold increases in $m$ cannot yet compare with an elevated $v$, in terms of facilitating an increase in $V_f$: as mentioned earlier, the regime in which $m\rightarrow cm\Rightarrow V_f\rightarrow cV_f$ applies is where $m\ll M$. In this dynamical regime, the mass-transformation might look like donning a 16-$oz$ glove to double the $m$ of the striker's hand before aiming a ``body shot" at the defender's abdomen $M$ -- hardly the impressive image of $kV_f$ as increased ``power."

On the other hand, the accessible ranges for incoming velocity values $v$ are also bounded~\cite{walilko2005biomechanics,stanley2018analysis}. Even amateur fighters do not tend to punch with speeds as slow as $4~\frac{m}{s}$, and it is exceedingly rare for most professionals to strike with hand speeds that are consistently in excess of
$10~\frac{m}{s}$. Although a Guinness World Record~\cite{guiness2013} cites one punch clocked at 45 mph ($>20~\frac{m}{s})$, the abundance of $10-14~\frac{m}{s}$ upper-limb strikes \emph{during combat} cannot be understated, especially without doubt on the ecological validity of the research settings in which such rapid strikes have been recorded~\cite{beranek2023force}: strikes can be evaluated on a basis of speed or efficacy in laboratory settings, but may not reflect how punches are thrown (or received) in the sport-combative context of the ``ring" or ``octagon". Bounding the
values for the incoming strike velocity $v$ means restricting those allowed values of its multiplicative factor -- used above to define $V_f\rightarrow bV$, for $v\rightarrow bv$ -- to a range like $b \in [0.5,1.75]$ if we take $v=8~\frac{m}{s}$ as the median value -- far tighter than that of the mass multiplier, which was, loosely, $c \in \left[1,6\right)$.

A still-loose but more realistic, theoretical lower-bound for punching speed, $v\sim 6~\frac{m}{s}$, together with upper bound denoting \emph{repeatable}, unremarkable, non-outstanding performance around $10~\frac{m}{s}$, would further restrict the span, or possible-improvement range, of incoming speeds to a factor of $\sim1.67$-fold from bottom- to top-end exemplars. Assuming a fighter began at the low end, and propagating the full speed increase to target velocity $V_f\rightarrow1.67V_f$, we would find an overall effect that would be much more commensurate with, and obtainable via, a technique- or training-induced increase in effective mass $m$ instead.

Thus, a UFC-sized fighter, initially incorporating some 5-10\% of their full bodyweight into their strikes, might be capable of achieving increases in the post-collision target velocity $V_f$ that were previously thought to be achievable exclusively through incrementing $v$. Since the factor by which $v$ can update, via training, saturates at something like a $b\sim 1.67$-times increase, ratcheting up the hand speed could theoretically be swapped for changing a $5kg$ incoming, effective mass from $m\rightarrow1.75m = 8.75kg$, for the \emph{same effect} on $V_f$ as the maximum-possible, training-induced speed increase, against an $M=70kg$ target. All that remains to be seen is whether any \emph{training-mediated} increases in effective mass~\cite{neto2007role} can approximate $c\geq 1.75$; even maximizing muscle-mass gain~\cite{chappell2018nutritional} saturates $c\leq 1.5$.

In Figures~\ref{fig:vf_60kgtarget_mhoriz}-\ref{fig:vf_60kgtarget_vhoriz}, we plot example curves indicating how both the incoming, effective mass $m$, and its initial strike velocity $v$, determine the post-collision velocity $V_f$ of the target, point mass $M$ under conditions of the elastic, one-dimensional collision described above. For $m$, we bound all plotted values to roughly $\leq80\%$ of the UFC Heavyweight delineation -- the highest cutoff for allowable UFC weight classes~\cite{unifiedrules2024}. Similarly, for the velocity input $v$, we extend boundaries slightly beyond the $\sim8-10~\frac{m}{s}$ range, so as to capture a reasonable extent of variation for each of boxing, karate, muay thai, and MMA generally~\cite{beranek2023force}.
% Consider supplementary scatter plots for meta data on striking velocities; Vf evaluations as rows in a table.

Taken all together, the results of this section so far demonstrate that -- for \emph{realistic} values of $m$, $M$, and $v$ -- expectations that tuning $v$ alone can achieve a substantial increase in $V_f$ would perpetuate an altogether inaccurate description of this collision phenomenon. Below, we will extend these arguments and call into question additional variables besides $V_f$ as outcome metrics for a ``powerful" punch; for now, we first comment on why $V_f$ is useful.

\section{Other Dynamic Variables, and Model Generalizations} 
\label{other_variables}

\subsection{Whiplash: Arguing for the Relevance of $V_f$}
\label{whiplash}

Since the value of $M$ is tacitly assumed to remain fixed during our collision, the total \emph{impulse} experienced by the (effective) target mass $M$ is maximized for maximum $V_f$:

\begin{eqnarray}
\argmax_{V_f}\left(J\right) = \argmax_{V_f}\left(M \int_0^{V_f} dv\right) 
\label{eq:8}.
\end{eqnarray}

% from which it is clear that $V_f$ 
% wehre the 

For a point mass with a single, translational degree of freedom, the post-collision velocity $V_f$ is a natural ``outcome" variable for reconciling the initial conditions of the impact -- yet, we contend, one that may be overlooked in combat sports applications. Once additional degrees of freedom (especially in the form of incorporating multiple mass components) and details like material inelasticities are taken into account, it would seem that the usefulness of $V_f$ for assessing, e.g., strike damage potential, would vanish. On the contrary -- at least where incoming energies (see Section Za) are not absorbed, to grand degrees, by the target or ancillary masses tightly coupled to $M$ (such as a mounting bracket, or even the ground) -- the post-collision velocity $V_f$ offers useful information about the mass ratio $u\equiv\frac{m}{M}$, and remains a measurable datum. % Q&A: is this information unique?

As outlined above, $V_f$ will assume a substantial value ($v\leq V_f\leq2v$, for $m \geq M$) only when the initial strike momentum $mv$ is itself transferred in such a quantity that $M$ gains in speed appreciably. Although this transfer is mediated by both $m$ and $v$ independently, with different associated transfer functions, we have seen that an increase in either of these input variables can effect changes in $V_f$ on an order of magnitude similar to those realizable by the other, once we take into account realistic ranges of variation in applications to real human combat.

What differs are their effects on $v_f$, the continuation or recoil velocity of the initially-moving, striking mass $m$: solely in the instances where $m$ experiences waning recoil, or ceases its forward motion
along its initial path of travel can we logically assume that the incoming and stationary effective masses are essentially matched ($m\sim M$). When this happens, the full momentum $mv$ is assumed into the target. Whenever $m<M$, we can go a step further and take the initial value of $v$ as a bounding value for $V_f$, i.e., an upper limit, such that $V_f<v$. In this case, as for two human bodies colliding in such a way that their effective mass fractions are comparable, the value for $V_f$ can be ``read off" as a direct, experimental proxy for $m$ (indeed, a recoil speed $||{v_f}||>0$ for $m$ can also be used to infer the mass ratio $u\equiv\frac{m}{M}$, and therefore $m$, when the target mass $M$ is known; we discuss the utility of this inference elsewhere, and focus here on the relationship with $V_f$).

Now consider a simple extension of the model espoused above, wherein the mass $M$ is comprised of not one, but two, distinct components. A sample configuration for the %two
components is shown below, in Fig.~\ref{fig:stickmodel}, where some portion ($<30$\%) resides on top of a rigid, massless stick, and the ($\sim70$\%) majority lies below, affixed to the bottom of a second, identical stick. The sticks are connected by a ``hinge" joint, and oriented vertically prior to impact. When the incoming mass $m$ strikes the target, it does so near the vertical center -- the halfway point -- between the components of $M$, and not necessarily at its center of mass. According to the laws of inertia and conservation of angular momentum, the upper component of $M$ ultimately tends to acquire a linear, tangential speed greater than the magnitude of the velocity of the center of mass of $M$. This effect is a primitive model for \emph{whiplash}.

In reality, the combined segmental weight of head/neck ($<8$\%) relative to the ``chest" constituents of the thorax ($\leq 20$\%) for a stationary, defending fighter -- as well as the significantly more-superior anatomical location of the skull, compared with the hinge joint in the ball-and-stick diagram of Fig.~\ref{fig:stickmodel} -- will tend to ensure an even greater tangential velocity accentuation, and both a higher linear \emph{and} rotational acceleration for the defender's brain, than would be calculated within this crude approximation.

\begin{figure}[ht]
\includegraphics[width=0.5\textwidth]{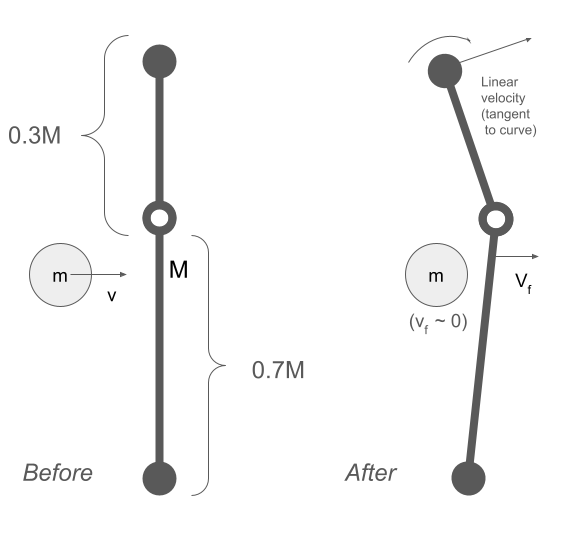}
\caption{\label{fig:stickmodel} Graphical illustration of the parameters involved in a collision with a simple, ball-and-stick model of a fighter's body. The incoming, effective mass $m$ strikes below the hinge that divided the top $30\%$ of the target mass $M \sim m$ from the bottom $70\%$; while not drawn precisely to scale, the value of $V_f$ approximates $v$, whereas the uppermost component of $M$ moves in an clockwise arc at such an angular speed that that its linearized, tangential velocity can ultimately exceed $V_f$.}
\end{figure}

The rapid \emph{deceleration} that then inevitably occurs as the brain organ rebounds off the skull in a secondary, internal collision, would itself qualify as a primary risk factor for whiplash-induced concussion, equally as a positive acceleration, if it exceeds its respective threshold~\cite{towns2026linear}. Where acceleration values cannot be obtained directly for comparisons, the full-body (or now, the center-of-mass) velocity, $V_f$, immediately post-collision, gains usefulness as a proxy that can be extrapolated to a representative radius to the brain, given a model like the one illustrated earlier. For an incoming effective strike mass of $m\sim 5kg$, comparable to the mass of a human head, we can directly verify that $V_f \sim v$; for both smaller and larger incoming masses, we can reference Eq.~\ref{eq:4} to estimate $m$ from $V_f$.

In practice, any real material -- including human body tissue -- will ``scatter" the initial momentum of a striking implement into various directions that are not necessarily pointed along the initial vector direction of $v$, among the movable components of $M$; convert some of the initial kinetic energy $\frac{1}{2}mv^2$ to elastic potential energy in the springlike components of $M$; and dissipate kinetic energy (i.e., \emph{inelastic} collisions) into forms like heat and sound.

Accordingly, any ``absorption" of energy into sensitive body parts ~\cite{natale2026_wp2} that diminishes the value of $V_f$ and therefore weakens the ability to use center-of-mass velocity as a proxy for $mv$ would seem to threaten the hypothesized relationship between $V_f$ and $m$. Given that real materials differ in their attenuating (i.e., impact energy-absorbing) properties, however, it is in theory possible to select materials that selectively extinguish the $V_f$-boosting effect of $v$ and instead focus on the variation of $V_f$ with $m$~\cite{liu2015compressive}. Consequently, it continues to be possible -- and possibly advisable, under certain circumstances -- to use the presence and severity of a whiplash effect on the target $M$ as a proxy for $m$, in practice. Specifically, a strong, clear, impulsive whiplash (as opposed to a more brief \emph{jolt} or vibration) has been used in this manner for the order-of-magnitude assessment of $m$ in the context of video analysis~\cite{natale2026_wp2}. Taken to its extreme end, the logic within Eq.~\ref{eq:3}-\ref{eq:4} implies that simultaneous observation of minimal recoil of $m$ with \emph{any} marked acceleration of the target $M$ -- from unadulterated whiplash down to plain, gross motion -- is a sign that two initially-moving and initially-stationary masses are matched. This fact offers a practical, diagnostic criterion that separates a coach's ocular estimation of $m$ from that of $v$ (the latter being more amenable to, and requiring, quantitative data for high-fidelity inference).

\subsection{Scaling of the Kinetic Energy, in Combat-Relevant Regimes} 
\label{scaling_ke}

In the absence of external forces, momentum shall be conserved in all the collision variations discussed here. In addition, the presumed elasticity of the collisions ensures that ``kinetic energy," in the functional form $K=\frac{1}{2}mv^2$, summed across both the system components, will also be preserved, in total value. The kinetic energy is sometimes studied 
as a loose proxy for target \emph{damage} in collisions that admit some inelasticity~\cite{wasik2009chosen,del2022reliability}, given that its dissipation (e.g, conversion to heat) and availability to induce gross motion (or, increase potential energy in the target), per the relationships given by the so-called Work-Energy Theorem~\cite{ling2016university} -- more directly  than momentum -- track the amount of mechanical energy ``dumped" into a target $M$.

Arguments for direct relationships between proficiency in striking and a high initial velocity in the effective mass $m$ tend to exploit the previous functional form for kinetic energy $K$ of the incoming mass, to illustrate the explicit relevance of the value of $v$ to increasing strike energy.
In particular, the incoming velocity is popularly designated as a somewhat unique determinant of striking ``power" (in the nontechnical, anecdotal sense of the word), due to the fact that $v$ enters as a \emph{square} in the above expression; the only other variable, $m$, appears merely as a term to the \emph{first} power. Strictly speaking, this scaling logic leads back to an argument for the dominance of $v$, over $m$, in determining the energy of motion -- at least, if both were to increase, respectively, by similar factors.

To wit, consider a strike with initial velocity $v=8\frac{m}{s}$, with an effective mass of $m=5kg$ (comprising, for the sake of argument, a fighter's fist, lower arm, upper arm; perhaps some additional connective tissue encompassing the glenohumeral joint attachment) -- in other words, its initial momentum measures as $mv=40Ns$. As an alternative configuration, let us \emph{quadruple} the incoming mass such that $m=20kg$ -- but also slow the incoming velocity to a mere $v=2\frac{m}{s}$, to preserve that nominal momentum value of $40Ns$. Such a transformation, $m\rightarrow cm$, leads to an effective mass more equivalent the fighter's entire leg than a single arm for this case, $c=4$ but, combined with the concurrent decrease in the magnitude of the velocity, quadrisects the kinetic energy value ($K\rightarrow\frac{1}{4}K$). Another approach to building intuition about the relevance of the velocity in determining the kinetic energy value would be to ask what the value of $m$ must be to match the energy $\frac{1}{2}\left(5kg\right)\left(8\frac{m}{s}\right)^2=160J$ of the first configuration above; at $v=2\frac{m}{s}$, this case would require a staggering $m=80kg$, a value certainly out of reach, short of literal ``full-body" strikes. Even a moderate increase in the velocity, though, back to $v=4\frac{m}{s}$, would restore the kinetic energy value to $160J$, if endowed to the aforementioned, $20kg$ mass.

Despite the transparency of the expression $\frac{1}{2}mv^2$, the further claim that the magnitude of the strike velocity $v$ necessarily has an unambiguously \emph{greater relevance} than $m$ in determining the extent of ``target damage" is not as well supported. This is for two reasons. First, in practice, the circumstances under which a given strike \emph{lands} upon its target (including material interactions) may explain a percentage of variability in $V_f$ at least as large as that of the ``initial conditions," or dynamical variable values that set up the collision problem, do. Second, dynamical bounds (introduced Section~\ref{insufficiency_v}) can be used to establish how the previously discussed limitations on the accessible value of $V_f$ for human combat  -- insofar as strike impacts can be modeled as 1D, elastic collisions -- in turn limit the extent to which target kinetic energies, post-collision, can scale with the incoming strike velocity for across different values of $m$ (see Fig.~\ref{fig:KE_60kgtarget_mhoriz}-\ref{fig:KE_60kgtarget_vhoriz}). Perhaps counterintuitively, the variation of $K$ with the incoming, effective strike of mass $m$ can be as substantial as its variation with $v$ (like $V_f$).

As a concrete example, for a target of mass $M=60kg$, there is less than a $3$-fold increase in $K$ associated with increasing $v$ from $6\rightarrow10\frac{m}{s}$ -- i.e., the $b\sim1.67$ discussed earlier for $V_f$, itself. Naturally, this relationship holds steadily for all fixed $m$ values between the $5kg$ baseline value, alluded-to earlier as the approximate upper-limb mass for a Lightweight fighter, and its quadruple ($c=4$), $20kg$, representing a practically conceivable upgrade that extends to approximately half of the torso / ``trunk" mass as well. Meanwhile, the kinetic energy increase achieved by incrementing the mass $m$ by such an amount, for \emph{fixed strike speeds} $v\in \left[6,10\right]\frac{m}{s}$, spans a full order of magnitude (in the language of the previous section, $k\sim10$). This is because the effects of increasing mass, by any multiple $c$, becomes stronger for large, combat-relevant speeds.

Conversely, it must be acknowledged that incrementing the incoming strike velocity, within that aforementioned, realistic range $v\in \left[6,10\right]\frac{m}{s}$, still effects a non-negligible, nearly- but still super-linear increase in $K$ once incoming effective mass values $m$ rise to $\sim20kg$ or higher (below this amount of incoming mass, slopes are small, and even more linear; beyond $v=10\frac{m}{s}$, $K$ grows at a measurably faster rate than a strictly linear slope). In addition, as for all the preceding discussions, there is nothing~\footnote{That is, nothing explicitly \emph{physical}; a few \emph{physiological} constraints may play a role in preventing some athletes from attaining the high-end speeds once their weights transcend a certain level, but even this pattern is not universal; moreover, it is recommended to pursue increases in $m$ through technique modifications, rather than, e.g., augmenting the size of the striking limb hypertrophically.} to stop a given athlete from pursuing trainable improvements in both $v$ and $m$ to maximize momentum or impact energy.

\subsection{Inelasticity in Real Punches}
\label{inelasticity}

The authors of a prominent review on the relevance of the concept of \emph{effective mass} to striking performance in a combat sports context, specifically~\cite{lenetsky2015effective}, have admonished against misleading oversimplifications that may arise via modeling an athlete's striking implement as a ``uniformly shaped block of mass"; a related criticism, found therein, is against
elasticity assumptions in striking models~\cite{lenetsky2015effective}.

It might be due to fundamental misunderstandings of the mechanics -- perhaps, equally likely, due to differences in emphasis, with a focus on the striking mass $m$ and less on the target mass $M$, as per our nomenclature -- that the authors in question contend that ``the greater the rigidity of the impacting mass"  \emph{and, thus}, ``the less elastic a collision, the greater the momentum imparted into the target or opponent." We have so far avoided addressing implicit assumptions of our springlike, 
\emph{barely deformable} striking mass, which -- if violated -- would indeed modify the ultimate values of $V_f$, and the change the amounts of kinetic energy imparted to the target mass $M$. Still, quotations 
like the above seem to betray ``abuse of terminology". To wit, even at the highest levels of sports-scientific research on effective mass in striking, similar-sounding terms may have been ``loaded" with multiple meanings that conflict with their standard usage in physics and engineering.

Conventionally, \emph{higher} degrees of elasticity convey less energy loss -- less heat dissipation, upon target deformation, and generally \emph{more} transfer of momentum from the initially-moving, effective strike mass $m$ to the target $M$. %Stand in contradiction...!

Nevertheless, basic discussions put forth in Ref.~\cite{lenetsky2015effective} on the need to define an ``effective" mass (as some actively-participating fraction of the full body mass) does capture the idea of incompletely-coupled body segments correctly -- additionally, their notes on ``follow-through" do fairly accurately, if imprecisely, articulate a notion of how elite fighters can contribute additional momentum in the short time frames that `define' a human striking collision~\cite{natale202610kfollowthrough}.

What might be found to be lacking, or even misleading, is the erroneous expression that \emph{inelastic} deformability in the striking implement (i.e., a net \emph{loss} of energy upon returning to its pre-collision shape, or configuration) could ever \emph{facilitate} momentum transfer to the intended target.

Furthermore, no quantitative support is given in favor of any specific approaches to modeling ``an actual strike," as opposed to ``ballistic modelling of effective mass" [\emph{sic}]. In this article, we study the simplest possible valid model. As a matter of fact, the most natural place for \emph{inelasticity} to fit into the simplistic model espoused here is an energy-absorbing deformation in the target mass $M$, else mutual deformation shared between the striking mass $m$ and the target. In this situation, the post-collision velocity of the target will be given not by Eq.~\ref{eq:4}, but by 

\begin{eqnarray}
V_{f,~inelastic} ~=~ \left(\frac{m}{m+M}\right)v%
\label{eq:10},
\end{eqnarray}

a factor of two smaller than our result for the perfectly elastic case. It should be clear that each of the preceding arguments can be modified, straightforwardly, by making such replacements to gain insight into the order-of-magnitude effects of adjusting $m$ and $v$, respectively (as well as comparing these effects, as before), on $V_f$. Moreover, it can be shown that even perfectly elastic collisions can be modeled as consisting of an explicit, and perfectly, inelastic phase~\cite{kidspaperCUNY}, which obeys the conservation laws we invoked, at the inception of this article, to write Eq.~\ref{eq:1}-\ref{eq:2}.

\subsection{A Note about ``Power" in Striking}
\label{power_conventional}

Continuing momentarily on the theme of ``abusing" of terminology in martial arts and combat sports -- that is, assigning of multiple meanings, across the diverse physical and non-technical lexicons, to the same words -- it is worthwhile to note here that there is no universally recognized way to measure a unique quantity called `punching power.' Like \emph{effective mass}, the usage and very meaning of this term varies across different combat communities, changing according to context within a given community, and it is unlikely that the na\"ive application of definitions from physics or engineering (e.g., mechanical work done, or energy expended, per unit time) would address what is intended in mainstream fighting references. Even where the standard physics jargon does apply, the concept of ``power" is itself is doubly-valued, possibly referring to the power generated during the acceleration of the effective striking mass $m$ until it reaches its starting velocity $v$ prior to a collision (i.e., the initial effort expended by the striking fighter), or to some time derivative of particular changes in energy associated with the ensuing impact.

This is not the case for all scientifically-derived terms, and other ``physically"-recognized, dynamical quantities. Contemporary sports science has no issue characterizing velocities, accelerations~\cite{ishac2021evaluating,mosler2024higher}, and impulses~\cite{walilko2005biomechanics,adamec2021biomechanical,beranek2022performance,mosler2024higher} in a relatively unambiguous way. Even ``momentum," once a `term-of-art,' has filtered strongly into the growing vernacular of recent studies on fighters, with trainers often using this word correctly, with regard to coaching the buildup to an impact event. The phrase ``strength," like ``force," suffers from ambiguity stemming from a lack of appreciation of the continuous-time nature of Newtonian forces, although certain studies have become more specific and adamant in mentioning and measuring ``peak" forces. For further discussions on how \emph{force} relates numerically to the treatment of variables like $V_f$ above, and to the rest of the dynamical quantities plotted throughout the following Section -- as well as a more thorough critique and an alternative method of measuring \emph{effective mass} -- we direct the reader to the ongoing work~\cite{natale2force,natale2026a}.

\subsection{Target Acceleration \& Special Case of $m\gg M$}
\label{concussionthresh}

In Section III, as above, we will focus predominantly on the regime in which $m<M$ -- in practical terms, this represents a strike (e.g., of effective mass $m\sim5kg$) to the most massive, core components of the abdomen or thorax of a defending fighter's body (say, $M\sim60kg$). By virtue of the Hill Force-Velocity curve~\cite{hill1938heat}, we expect that such a small striking implement -- be it the offensive fighter's hand, elbow, etc. -- should be of ``small enough" mass, compared to the striker's body mass and muscular-energy resources, that it can be accelerated to an appreciable $v$ (commensurate with $6-10\frac{m}{s}$ ballpark advertised earlier).

In this subsection, we momentarily turn our attention to the case(s) $m\geq M$; consider a swap of values, so that $m\sim60kg$ and $M\sim 5kg$, as one imaginative example.

Assuming that the stationary mass $M$ is relatively non-dissipating, such that all our elasticity assumptions hold, values of $V_f$ and $k$ -- as dictated by Eq.~\ref{eq:4} and Eq.~\ref{eq:7b} -- can differ rather significantly from those explored in other examples provided throughout the previous sections. Given $u=\frac{5}{60}\approx.083$, such a strike would almost place us in the $m\gg M$ regime, where $V_f\rightarrow2v$. It is trivial to calculate the target's acceleration, from rest, as $a_M=\frac{\Delta v}{\Delta t}=\frac{V_f}{t_{coll}}$, where the ``collision time" $t_{coll}$ measures the duration of appreciable physical contact between $m$ and $M$. While we discuss the direct measurement of contact times elsewhere~\cite{natale2026b}, one approximation might be $5-30ms$~\cite{lenetsky2015effective} for a \emph{gloved} punch to the \emph{head} of a defending, a target which tends to satisfy $m\sim5kg$~\cite{plagenhoef1983anatomical}, regardless of weight class.

In conjunction with earlier notes (Sec.~\ref{whiplash}) on brain acceleration and deceleration -- which should lie in a range \emph{below} approximately $a_M\in\left[70,100\right]g$ to avoid concussion %with a 
with a significant degree of probability~\cite{broglio2010biomechanical,towns2026linear,tiernan2020concussion} -- the relative constancy of $k$ in the limit $u\equiv\frac{m}{M}\rightarrow \infty$ implies that the chances of a strike-induced ``knockout" turn out to be somewhat independent of the exact value of $m$ here. More precisely stated, there should exist \emph{i)} an order-of-magnitude range for the values of $m$ that can serve as a kind of ``minimum effective dose," above which the ability of any strike to induce a concussion depends solely on a sufficient initial velocity $v$, for a given collision time ($t_{coll}$ tends to be characteristic of the interacting materials). % We assert m in another publication; here we mean spring.

Interestingly, once the effective mass $m$ reaches some substantial value, around $40kg$, target accelerations will exceed $70g$ for any strike with a velocity $v$ (immediately pre-collision) greater than roughly $8\frac{m}{s}$, for an $M=5kg$ collision that endures for $t_{coll}\sim20ms$. Although higher velocities $v$ can render higher $V_f$ values, it is found that even collisions for which $m\sim$ can induce similar target accelerations for more modest initial strike speeds, as low as $v\geq6\frac{m}{s}$, if the collision time is reduced to $t_{coll}=12ms$. 

Finally, extremely short-time collisions, lasting $\sim7ms$, will reach the threshold of $\sim70g\approx 686\frac{m}{s}$ for nearly any combination of $m\sim5kg$ with any of the aforementioned values of $v$ -- meaning that, from the perspective of the target acceleration, strict adherence to the limit $u\rightarrow 
\infty$ is not even necessary to risk inducing a concussion: starting with $m$ essentially commensurate with $M$, nearly any mass striking with sufficiently high $v$ is able to accelerate a target as small as $5kg$ at rates unsafe for human brains.

Conversely, a degree of safety-assurance can be realized by ``slowing" interactions until they last $t_{coll}\geq 40ms$~\cite{lee2014striking}, in which case our analysis suggests that $70g$ accelerations would not be encountered short of $60kg$ moving at $14\frac{m}{s}$, or $40kg$ moving at $16\frac{m}{s}$ -- above expectations for the UFC.

\section{Numerical Evaluations \& Discussion of Principal Results}
\label{numerical}

In this section, we analyze scaling behaviors for various quantities ($V_f$; $k$; kinetic energy) that we have explored above, characterizing outcomes for generic martial arts or combat sports-style strikes, to the extent that strikes can be modeled as one-dimensional, elastic collisions between point masses. To do so, we must decide on and designate reasonable baselines for comparison. When necessary, we default to a modeling results for an archetypal $m\sim70kg$ fighter -- sitting, therefore, at the upper limit of the UFC Lightweight class -- who strikes with $7\%$ of full body mass, for $m=4.9kg\approx 5kg$. Analogously, we declare the initial, pre-collision strike speed as a conservative
$v=\left(8\pm 2\right)\frac{m}{s}$ when investigating variations with $v$, for any fixed mass.

Unless otherwise noted herein, we focus our numerical evaluations on a target of $M=60kg$, representing close to the extent of mass that any given strike can encounter during combat (i.e, not merely the core/trunk, but most of the body, for similarly-sized opponents). Studying this amount of ``dead weight" allow us to map quantitative outcomes for the worst-case scenario, the combat equivalent of ``hitting a wall," or the case $m\ll M$: it encapsulates all but the most inferior (relative to the small-mass head) and ground-connected body segments of an opposing $70kg$ fighter, about any which rigid-body rotation and translational readjustment upon impact would occur.

% \subsection{Post-Collision Velocity $V_f$}
% \label{numerical_vf}

\begin{figure}[H]
\includegraphics[width=0.5\textwidth]{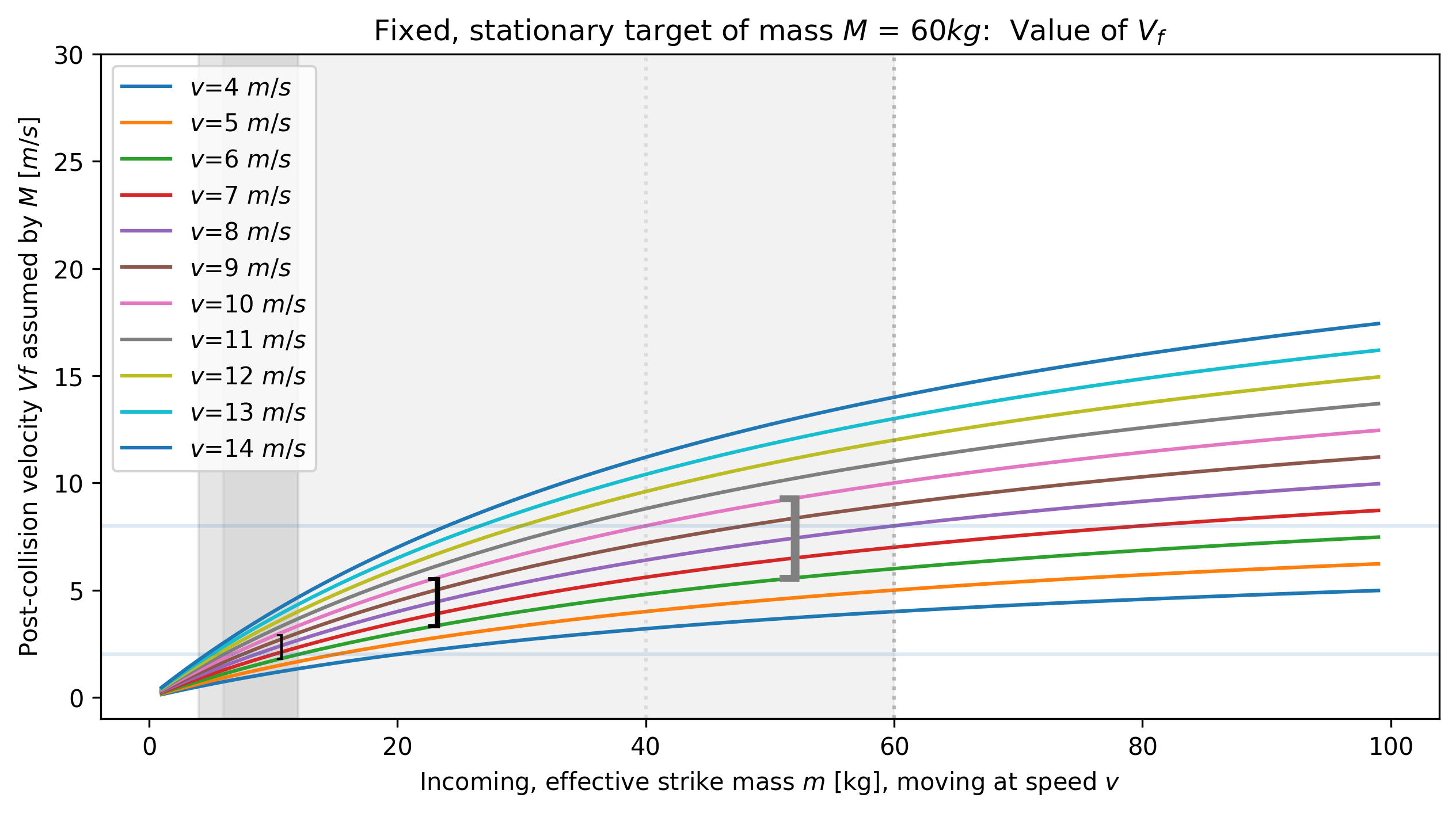}
\caption{\label{fig:vf_60kgtarget_mhoriz} The variation in post-collision target velocity $V_f$ with the incoming mass $m$ follows a consistent functional pattern, for which the strike velocity $v$ serves as a ``shape" parameter.} 
\end{figure}

We plot, in Fig.~\ref{fig:vf_60kgtarget_mhoriz}, variations in post-collision velocity for a one-dimensional collision, using Eq.~\ref{eq:4}, against both values of the incoming, effective mass $m$ of the striking implement (on the independent axis) and the initial, pre-collision speed $v$ of that moving mass (colored curves).

While we allow included ranges for these two variables to extend considerably beyond the lower and upper limits established in Section I above, we highlight their reasonable extents of variation as follows. For the effective mass $m$, by definition a fraction of the striker's total body mass, we first shade values between $4kg$ and $12kg$ in gray. This interval serves throughout as an ``extended baseline" region for $m$,  beginning at $\approx 6\%$ of the full, fighting body mass for a $70kg$ striker -- a low estimate for the proportion of that body mass attributable to a single upper limb -- and ending exactly three times higher (i.e., $cm$, with $c=3$). As we remarked in Sec.~\ref{insufficiency_v}, the ``low end" is representative of a fighter at the top of
the Lightweight~\cite{unifiedrules2024} division, striking with a quantity of effective mass that is consistent with previous literature. The upper end of this interval matches a more-optimistic estimate of $10\%$ body mass for the Heavyweight fighter maximum ($265lbs.$).

In a lighter shade of gray, we highlight effective masses up to $60kg$, representing up to $85\%$ of the full, $70kg$ body mass as one arbitrary and purely theoretical upper limit, for the time being, on the quantity of mass $m$ conceivably available to interact with a comparably-sized target, $M$. In doing so, we place a secondary dotted line at $m=40kg$ to represent $\frac{2}{3}$ of the full $70kg$, for reference; however, we also extend the horizontal axis all the way to a maximum of $100kg$, for purposes of comparison to the effective mass values $m$ that fighters \emph{heavier than those allowed to compete in the Lightweight division} might reasonably realize, if found able to commit their own, aforementioned $85\%$.

The velocity curves begin at a value of $4\frac{m}{s}$, perhaps an unreasonably ``slow" figure for ecologically-valid striking, and end at an almost equally ambitious $14\frac{m}{s}$. In order to render more salient the values in our declared range of $[6,10]\frac{m}{s}$, we include vertical brackets to illustrate how $V_f$ scales with changes in velocity $v$, for $m\in[10,20,50]~kg$.

The order of magnitude baseline value  of $V_f$ ($m=5kg$, at any $v$, is $\sim 1\frac{m}{s}$. For any fixed value of
$m$, increasing $v$ -- due to its linear, multiplicative effect on $V_f$ -- provides a constant, marginal gain (per unit of strike speed), to the ``post-collision" target velocity. By the same token, gains in $V_f$ afforded by specific increases in $m$ become enhanced at higher strike speeds $v$: the \emph{fold-increase} in $V_f$ attained by transformations $m\rightarrow cm$ remains the same regardless of the value of $v$, but the potential for \emph{absolute} gain in $V_f$ grows when the strike speed is elevated -- e.g., for $c=2$, from $+0.79\frac{m}{s}$ at $v=6\frac{m}{s}$ to $+1.32\frac{m}{s}$ when $v=10\frac{m}{s}$. If a reasonable limit for mass transformations is $c\leq 5$, the expected gain in target velocity $V_f$ between the same two strike speeds, $v=6\frac{m}{s}$ and $v=10\frac{m}{s}$ (the $b=1.67$-fold increase to which we alluded in Sec.~\ref{insufficiency_v}) is $\approx +2.35\frac{m}{s}$.

In general, moving rightward (i.e., $c>1$) in $m$, from our arbitrarily-chosen baseline value of $m=5kg$, for fixed $v$, leads to numerical increases in $V_f$ that can mimic or usurp the effect of increasing $v$. At $c=2$ (i.e., realizing an effective mass $m=10kg$ instead of the $5kg$ default), we expect to watch $V_f\rightarrow~\sim 1.8V_f$ irrespective of whether the strike was thrown at $v=6\frac{m}{s}$ or $10\frac{m}{s}$ -- a $\geq 0.79\frac{m}{s}$ gain, as mentioned above -- which notably exceeds the magnitude of target velocity increase achieved by increasing $v$, %explicitly 
alone, from $6$ to $10\frac{m}{s}$ at baseline $m=5kg$. For $c=3$, gains in $V_f$ lie between $+1.48\frac{m}{s}$ (at $v=6\frac{m}{s}$) to $2.46\frac{m}{s}$ ($v=10\frac{m}{s}$) -- either way, $2.6$-fold increase from the value corresponding to the baseline case and approaching the aforementioned $+2.35\frac{m}{s}$ gain gleaned by increasing $v$ near the proposed, interim upper limit of effective mass for our $70kg$ fighter. By $c=4$, $V_f$ gains already \emph{exceed} this value for $v\geq7\frac{m}{s}$.

Focusing on those three brackets at $m\in[10,20,50]~kg$  affords additional insights from combined, simultaneous increases in both $m$ and $v$. Varying $v$ from $6\frac{m}{s}\rightarrow10\frac{m}{s}$, at each of the three designated, special points on the $m$ axis, we find that
the ``post-collision," target velocity increases from a magnitude of $V_f\approx2.29\frac{m}{s}$ (arithmetic mean of the results for $v=6$ and $v=10\frac{m}{s}$) at $m=10kg$, to between $3$ and $5\frac{m}{s}$ at $m=20kg$; finally, between $V_f\approx5.45$ and $V_f\approx9.09\frac{m}{s}$ for $m=50kg$. Thus, in absolute terms, the target velocity \emph{could} conceivably increase in magnitude by some $3-7\frac{m}{s}$, i.e., triple (for fixed $v$ -- either $6$ or $10\frac{m}{s}$), over the range $m=10kg\rightarrow20kg\rightarrow50kg$. We take this, in the context of the previous analyses, as consistent with our interpretation that \emph{larger mass values will amplify the effects of increasing} $v$, and not the other way around, in the dynamical the ranges that are theoretically accessible to our archetypal fighter. Were combat sports fought in a range such that $m\in[60,100]$, the velocity effect would be the strongest influence on $V_f$ for our $M=60kg$ target.

Given segmental weight distribution data for human bodies~\cite{plagenhoef1983anatomical,zatsiorsky1983mass}, we hypothesize that $m$ for any upper-limb strike, based on current techniques and sport-combative constraints are limited to increases that satisfy $c\leq 5$ -- a striking limb, supported by approximately half the torso; $c=10$, then, represents a truly gargantuan effective mass augmentation from a $5kg$ baseline, and should, arguably, be excluded from the present analysis. To the extent that $c=4$ could, in principle, be achieved via training~\cite{neto2007role} or changing technique~\cite{kacprzak2025biomechanics}, our results here suggest that increases in the effective mass of an upper-limb strikes can certainly equivocate with, or even outrun, improvements driven purely by increasing strike speed at baseline $m$.

\begin{figure}[H]
\includegraphics[width=0.5\textwidth]{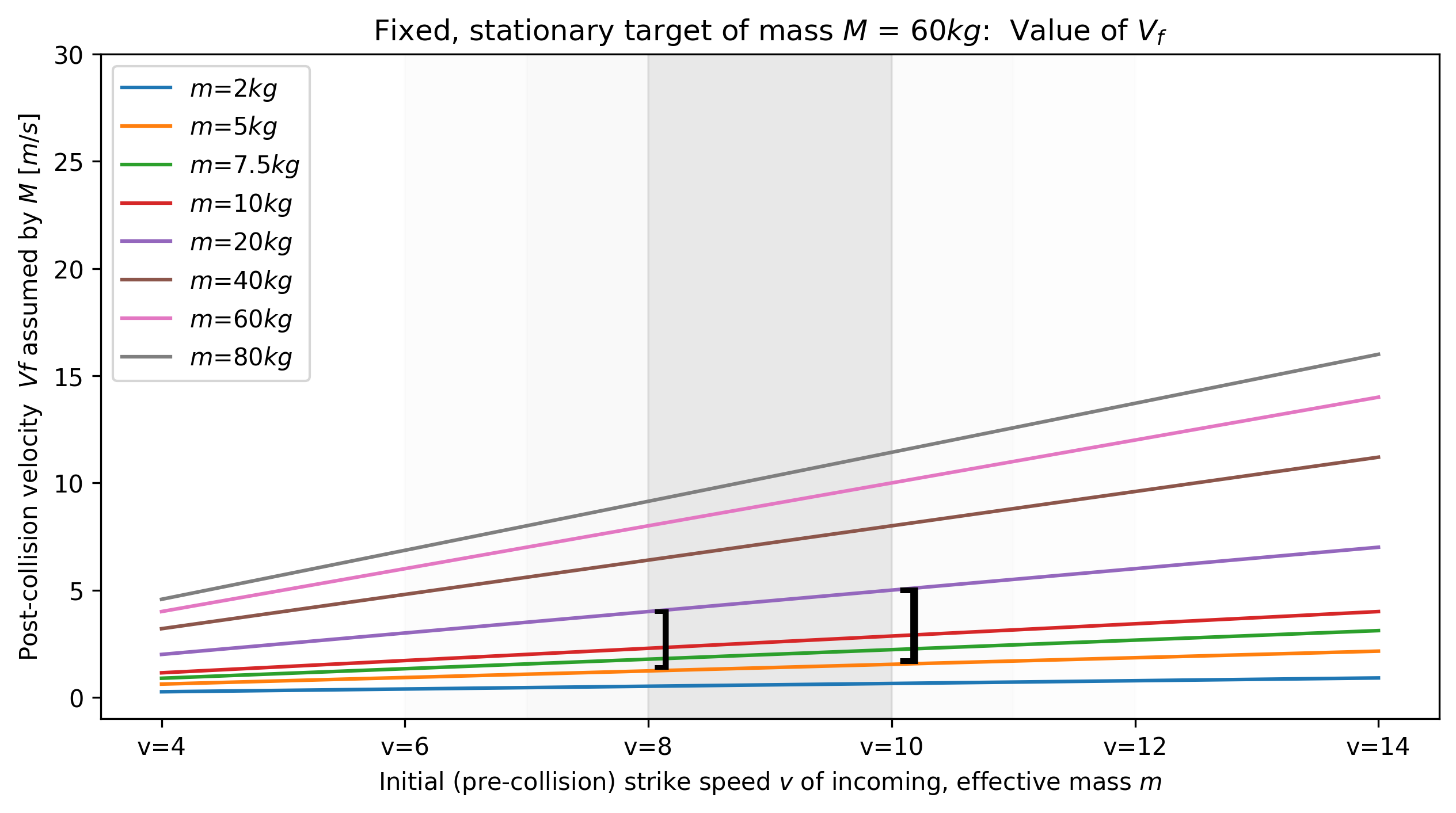}
\caption{\label{fig:vf_60kgtarget_vhoriz} Swapping the horizontal and vertical axes of Fig.~\ref{fig:vf_60kgtarget_mhoriz} may render more salient both \emph{a)} underwhelming response of $V_f$ to positive changes in $v$, in comparison with the almost-equally strong effects associated with increasing $m$ -- within a reasonable range, up to $\approx20kg$ -- as well as \emph{b)} the inutility of increasing $m$ to a value no larger than a $c=2$ multiple of its baseline, $5kg$ value. At $m\leq10$ (i.e., $c\leq2$), $v$ still dominates 
as an independent effect on the strike ``outcome" value of $V_f$.}
\end{figure}

This takeaway might be reinforced by simply swapping the horizontal and vertical axes of Fig.~\ref{fig:vf_60kgtarget_mhoriz}, the basis for 
creating Fig.~\ref{fig:vf_60kgtarget_vhoriz}.
Although only effective masses $m\geq20kg$ or so produce dramatic effects on the value of $V_f$ (i.e., for all lower values of $m$, the effect is comparatively smaller, and marginally near-constant), the fact that $m\sim20kg$ is close to saturating the hypothesized upper limit for mass (i.e., based on widespread understanding) can bolster
our argument that $V_f$ depends non-exclusively on $v$, and perhaps more strongly on $m$ in the dynamical ranges of importance for real fighters. Comparing the middle bracket in Fig.~\ref{fig:vf_60kgtarget_mhoriz} directly with those drawn at $v=8$ and $v=10\frac{m}{s}$ in Fig.~\ref{fig:vf_60kgtarget_vhoriz}, we can make use of both, matching $V_f$ scales to infer that mass transformations $c\in[4,5]$ produce similar gains in $V_f$, and similar ultimate $V_f$ values, as $v\rightarrow bv$.

At the mass values $m<20kg$ -- those more consistent with current literature -- increasing $v$ by quantities that have combat-relevant meaning, even when starting from a more realistic value of $8\frac{m}{s}$ instead of $6\frac{m}{s}$ (and reaching all the way to $12-14\frac{m}{s}$!) for professional fighters, affects $V_f$ predominantly by acting upon the platform provided, and reinforced, by $m$. At strike velocities of $8$-$10\frac{m}{s}$, the absolute gains in $V_f$ due to increases in the effective strike mass are substantial enough to recommend pursuing the theoretically-plausible transformations $c\in[2,5)$, by whichever, inventive training means are available~\cite{natale2026_shukjpotentialM}.

\begin{figure}[H]
\includegraphics[width=0.5\textwidth]{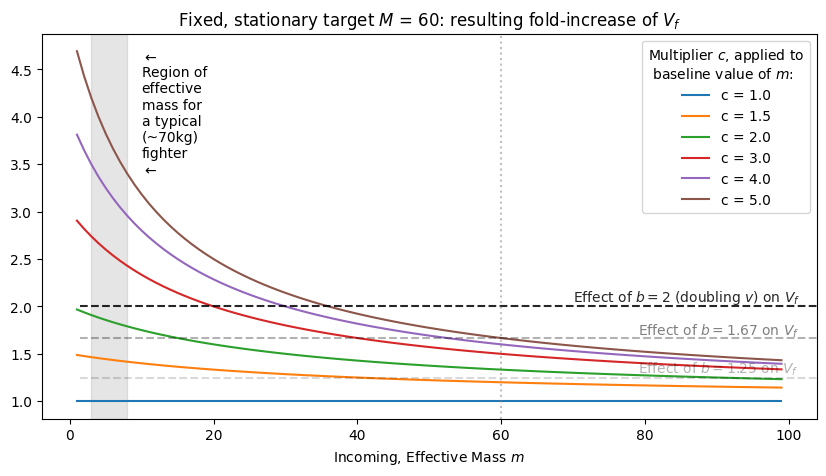}
\caption{\label{fig:k_manuallysaved} Considering $m<20kg$, there exists the distinct potential for the multiplicative transformations in the mass $m$ to cause more substantial, fold-changes $k$ in the value of $V_f$ than any of the reasonable increases in velocity ($b\in\lbrace{1.25,1.67,2\rbrace}$). Focusing on practically-realizable multiples $c\leq3$, specifically, the \emph{maximum} value of the incoming, effective mass $m$ that can be mustered for a \emph{starting} value of $20kg$ -- consistent with some $85\%$ of the complete body mass of our prototypical $70kg$ fighter -- is $60kg$, the conceivable ceiling for a full-body strike.}
\end{figure}

Even accounting for the smaller magnitude of $V_f$ values associated with ``near-baseline" values of $m$ and $v$, stories of the untapped, underappreciated potential associated with effective strike mass (i.e., as a mechanism of growth for the target velocity) are more clearly told in terms of achievable \emph{fold-increases}, rather than \emph{absolute gains}. A more explicit depiction of the dependence of $k$ on $m$, in comparison with values of $b$, can be found in Fig.~\ref{fig:k_manuallysaved}. The baseline range for $m$ is highlighted in gray, and the theoretical near-maximum of available effective mass (i.e., approximating the total body mass of the fighter) is set as $60kg$, for visual reference, via a vertical dotted line.

Again, it may be worth nothing that all the horizontal axis values $m>20kg$ in Fig.~\ref{fig:k_manuallysaved} represent values for the incoming strike mass $m$ that are not currently hypothesized to be accessible to our $70kg$ fighter; they may be accessible to strikers in heavier weight classes -- and, even then, possibly only those with specialized training; see comments below, on the feasibility of realizing $c\geq2.5$. Still, any large starting value for $m$ ``saturates" the effect of increasing the mass $m$ anyway, leading to substantially lower values of $k$; predicated on already-sizable, starting values of $V_f$, this could have been expected via theoretical analyses above (evaluating the $m\rightarrow\infty$ limit for Eq.~\ref{eq:7b}).

With regard to the opposite extreme, which concerns $m\leq 20$ and bears a contingently pessimistic, discouraging observation -- at least from the perspective of advancing our thesis, that effective mass $m$ matters, as much as $v$, for determining the outcome of agiven 
strike -- that the leftmost values of $k$ in Fig.~\ref{fig:k_manuallysaved} appear \emph{artificially} than the reference values of $b\in\lbrace{1.25,1.67,2\rbrace}$ against which they are juxtaposed, it is trivial to construct a counterargument. Where $k$ allegedly appears large \emph{solely} because the $V_f$ values corresponding to baseline effective mass values are themselves extraordinarily tiny, Figs.~\ref{fig:vf_60kgtarget_mhoriz}-\ref{fig:vf_60kgtarget_vhoriz} reveal that $V_f$ already assumes $\sim30\%$ at baseline, and at $10kg$ more than $50\%$, of its value at $m=20kg$ at $v\in\left[4,14\right]\frac{m}{s}$. Also, the scaling within even the leftmost gray, shaded region around our decided-upon baseline of $m=5kg$ is sufficient to demonstrate that $V_f$ has more growth potential with reasonable increases (scale-factor transformations) in $m$ than for the analogous, equally reasonable increases in $v$.

The proper utility of Fig.~\ref{fig:k_manuallysaved}, then, lies in \emph{i)} openly acknowledging the diminishing returns on increasing $m$, if $m$ were to start at another ``baseline" value already much higher than expected for some real fighter, striking with a single upper limb, and -- consequently, \emph{ii)} highlighting the fact that the region where one has the most to \emph{gain} -- in terms of $V_f$, target impulse, final target kinetic energy, etc. -- is precisely the region where $m$ has been measured to reside, for real combat. The \emph{opposite} is true of strike velocity: given its constant, multiplicative effect on $V_f$, $v$ would need to increase several-fold beyond the typical speeds seen in competition (from beginners to professionals) to outclass $m$ in the dynamical regions of interest.

Even so, something we have not yet addressed directly is the \emph{feasibility} of scaling the mass from a baseline value of $m=5kg$ to multiples of, say,  $c\geq2.5$ -- such that it represents a totality more akin to that of a leg than an arm (while still employing the upper limb as the striking implement) -- in practice. While it does not escape our notice that mass-transformations as drastic as for $c=5$ would imply that this athlete is striking with essentially the entirety of their available, bodily muscle mass -- or, equivalently, one upper limb reinforced by an overwhelming majority of the ``trunk" (i.e., chest, or thorax, as well as the abdomen and possibly part of the pelvis) -- % necessarily
strikes entailing such large proportions of bodily contribution have rarely, if ever, been reported in previous literature.

We hypothesize that such transformations $c$ are still a possibility in practice, not just by means of hurling one's body at an opponent in a manner that would compromise the striker's own balance and integrity of ``form" -- and furthermore posit that the reason for this absence of prior observation is due to a fundamental misunderstanding of how to recruit or incorporate body mass, effectively, into an upper-limb strike. Despite being out-of-scope for a full treatment in this article, we believe that perpetuation of this poor understanding has a concrete history, admitting a simple set of explanations regarding all those deviations of ideology that brought us to the modern paradigms for power generation. In a follow-up article~\cite{natale2force}, we discuss one alternative way to measure effective mass; we explore one promising mechanism -- and a trainable protocol -- for augmenting $m$, potentially to $c\geq3$, in ongoing work~\cite{natale2026_shukjpotentialM}.% on being done on a relatively unknown style of karate punch~\cite{natale2026_shukjpotentialM}.

Moving past the effects of increasing $m$ (and also $v$, by its respective, multiplicative factor) on the post-collision, target velocity $V_f$, we come to view the effects of both of these ``initial-condition" variables on the post-collision kinetic energy of the target -- representing a more general, all-encompassing proxy for strike \emph{damage potential}.

\begin{figure}[H]
\includegraphics[width=0.5\textwidth]{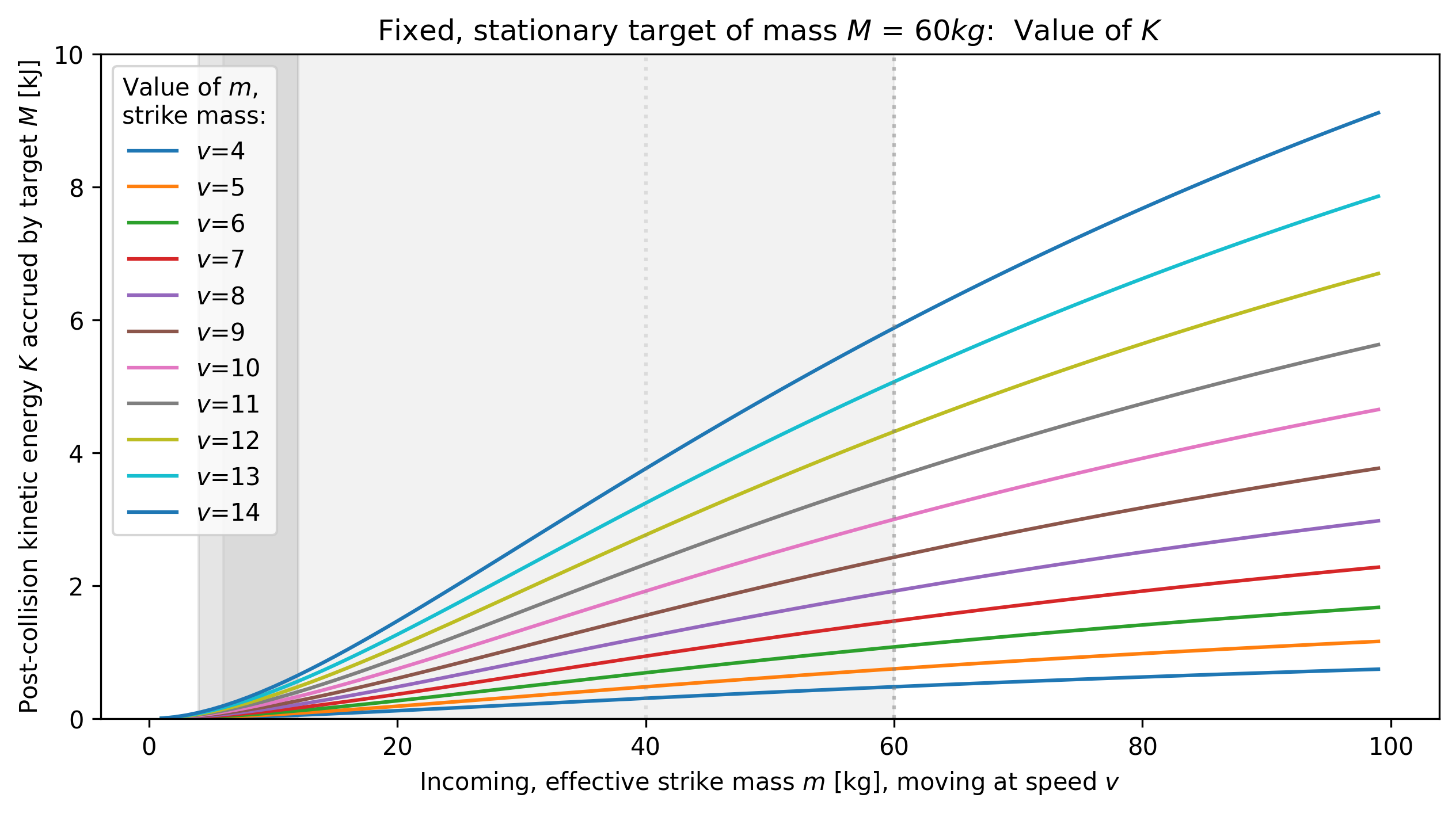}
\caption{\label{fig:KE_60kgtarget_mhoriz} Kinetic energy values for the target mass $M$ quickly as $m$ grows large, but most of the large values displayed here are inaccessible for real fighters. Nonetheless, the scaling between $m=5kg$ and $m=20kg$ at fixed $v$ remains sufficient to rival the reachable \emph{gains} in kinetic energy based on any velocity increase within our routinely-studied range of $v\in\left[6,10\right]\frac{m}{s}$.}
\end{figure}

Directing our attention, first, to the effects on $K$ that would be observed by varying the effective mass (Fig.~\ref{fig:KE_60kgtarget_mhoriz}), it is clear that the marginal gains in kinetic energy that are achieved, as a consequence of increasing $m$, depend heavily on the starting value of $v$ -- as before, the upshot in terms of one variable is modulated predominantly by the state of the other; somewhat in defiance of the ostensible separability, or factorization, of $m$ and $v$ in Eq.~\ref{eq:7}, this is due to the restricted ranges of both during combat. It is also here evident that this mutual dependence influences both the marginal growth and fold-increase of $K$ as $m\rightarrow20kg$ (i.e., $c=4$), or the hypothesized region of interest for practically-realizable effective mass values. At $v=6\frac{m}{s}$, the kinetic energy of the target would benefit from an increase of merely $\approx200J$; at $v=10\frac{m}{s}$, it could increase by $>700J$. Nevertheless, both of these marginal gains represent a full order-of-magnitude increase.

\begin{figure}[H]
\includegraphics[width=0.5\textwidth]{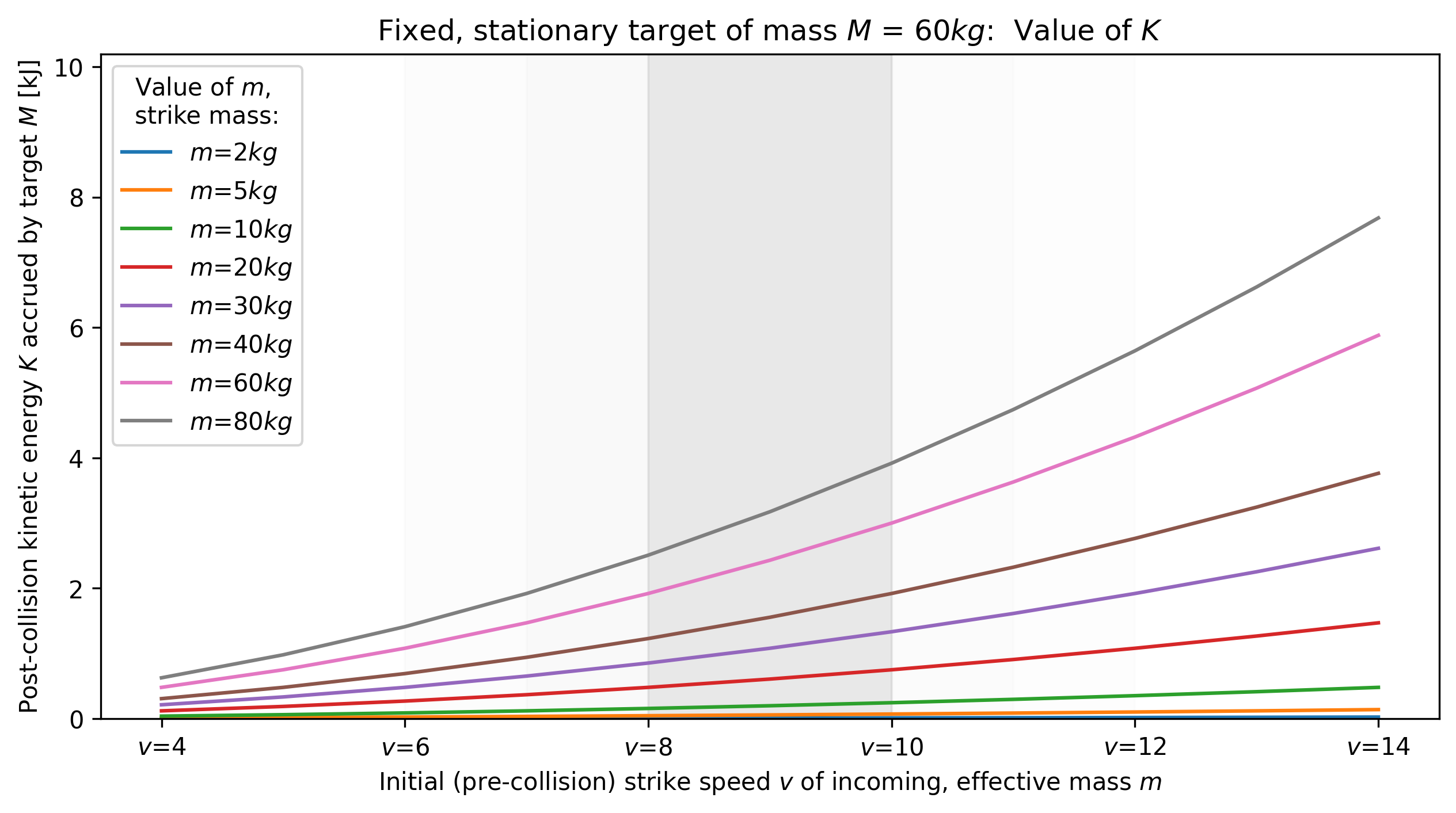}
\caption{\label{fig:KE_60kgtarget_vhoriz} Another presentation of a subset of the data in Fig.~\ref{fig:KE_60kgtarget_mhoriz} more readily demonstrates that the velocity-driven gains in the post-collision kinetic energy value for the target mass $M$ are small, relative to their theoretical, exponential growth potential, which comes into only once $v$ is rather larger than the magnitudes studied here). This is because we have restricted strike speeds to a range that is representative of those that can be consistently realized (e.g., during sanctioned combat) in practice. This result should continue to hold even for an observed shift of the median strike velocity value from $8\rightarrow10\frac{m}{s}$.}
\end{figure}

We can assess the rate of accruing marginal gains, for the (squared) velocity term, more transparently via the same ``hat trick" used to create Fig.~\ref{fig:vf_60kgtarget_vhoriz} from Fig.~\ref{fig:vf_60kgtarget_mhoriz}: switching out the independent axis in Fig.~\ref{fig:KE_60kgtarget_mhoriz}, we arrive at Fig.~\ref{fig:KE_60kgtarget_vhoriz}. 
In the regime where $m\leq20$ (i.e., $c\leq4$, from the assumed baseline value of $m=5kg$), the kinetic energy increases at a rate that just barely outruns a linear growth pattern. In contrast with the analysis example for $m$ above, the kinetic energy gain is all but negligible if $m$ remains at its baseline value of $5kg$ while $v$ increases from  $6$ to $10\frac{m}{s}$ ($+50J$, or a factor of $\approx2.78$), more substantial but commensurate with the results of the previous paragraph if $m$ already sits at $\approx20kg$ ($+500J$, or a factor of $\approx2.78$), and exceeding the results of the previous paragraph only if we assume both that $m$ sits near its upper, $m\approx20kg$ soft-bound for combat situations and $v$ is simultaneously taken from some conservative, low-end, amateur value to its practical maximum ($+600$, or a factor of $\approx2.25$, when from $v$ upgrades from $8$ to $12\frac{m}{s}$; $+1200J$, or a factor of $\approx5.44$, when transforming from $v=6\frac{m}{s}$ to $v=14\frac{m}{s}$).

\begin{figure}[h]
\includegraphics[width=0.5\textwidth]{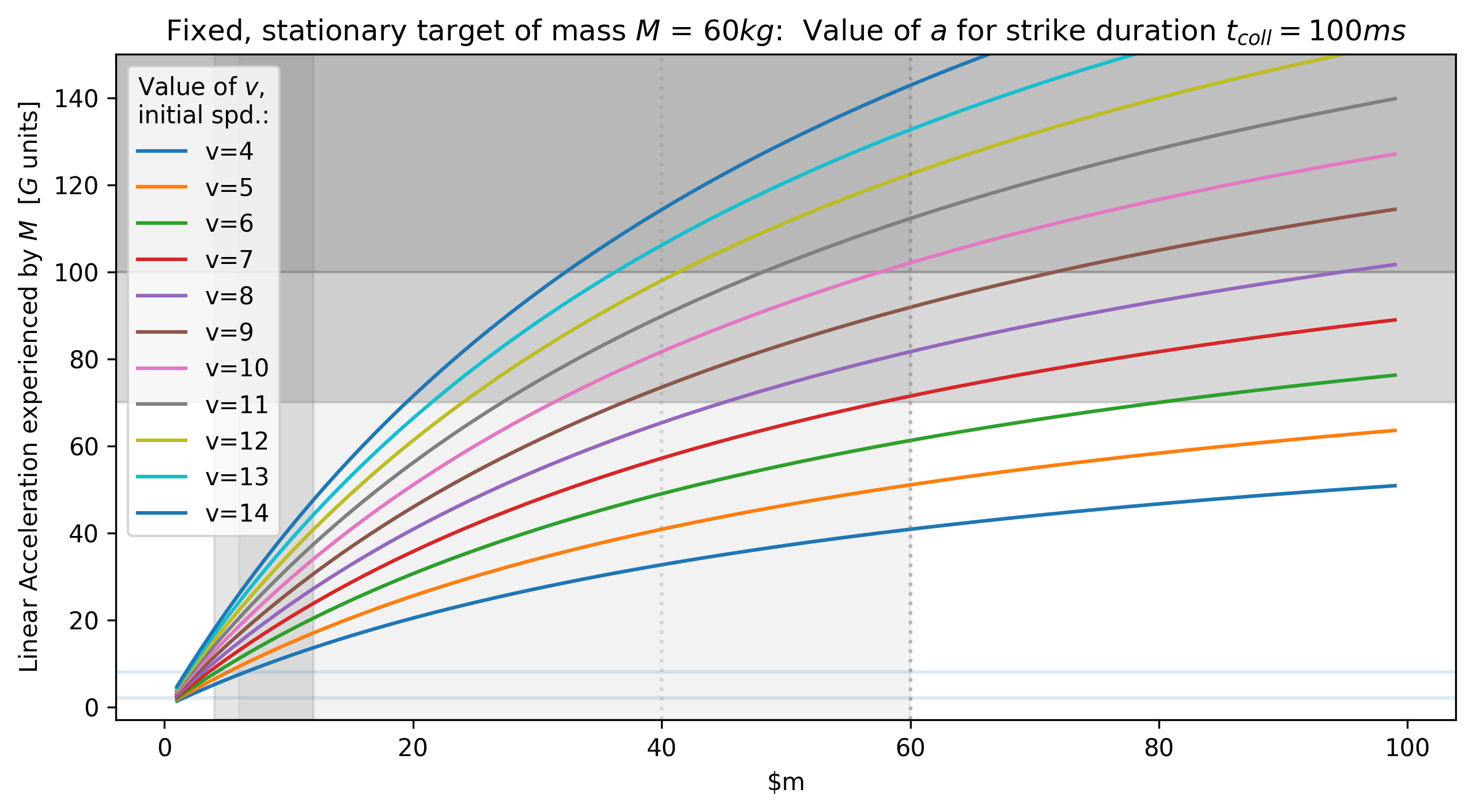}
\caption{\label{fig:accel_60kgtarget_mhoriz} Given a fixed collision time of $10^{-3}$ seconds, a target of mass $60kg$ will experience accelerations of order $100g\approx980\frac{m}{s}$ only for exceptionally large $m$ or $v$. In the principal dynamical region of interest for $m$, strike speed drives acceleration, exerting a multiplicative influence about as large as that associated with double or tripling the mass, but still insufficiently to reach explicitly dangerous magnitudes in $a$.}
\end{figure}

As our final set of numerical considerations here below, we plot select values for the linear acceleration $a=\frac{V_f}{t_{coll}}$ of the target mass $M$. We estimate $t_{coll}$ based on previous work that compares the time to peak force for MMA and boxing gloves~\cite{lee2014striking}; conservatively, assuming a symmetrical force-time curve to remain consistent with our model of elastic collisions, we double the mean value for the first $\sim 2000$ repetitions of $4oz$-glove impacts studied therein to arrive at $t_{coll}\approx10ms$. Highlighting both the dynamical regions of interest for effective mass values -- $m\sim\left[5,10\right]$, or $c\in \left[1,2.5\right)$ -- and likely-concussive impacts in progressively deeper shades of gray, we present these data in Fig.~\ref{fig:accel_60kgtarget_mhoriz} for our stationary target of $M=60kg$.

For any incoming strike velocities smaller in magnitude than $v=7$, we can interpret the set of curves in Fig V. as conveying that strikes consisting of $m\leq60kg$ tend to fail at passing $>70g\approx686\frac{m}{s^2}$ and entering the ``concussive" zone discussed in Sec.~\ref{concussionthresh}. For $m$ surrounding our baseline value of just $5kg$, even incoming strikes at $v=14\frac{m}{s}$ will tend to produce only about half the acceleration observed in various studies (e.g., football collisions that led to concussion diagnoses)~\cite{broglio2010biomechanical,broglio2012field,broglio2012high,hons2020measurement,tiernan2020concussion,tierney2024concussion}. %On the other hand, 
Nonetheless, strikes with at least $m=20$ (i.e., $c=4$, relative to the baseline value of $m=5kg$) threaten to exit the literature-established region of safety at such high $v$.

All our opening remarks on acceleration, however, still suffer from a fundamental inaccuracy of assumption: no human being's head has a mass of $60kg$. Thus, applying the aforementioned acceleration threshold as a proxy, or decision criterion, for neurological safety is problematic. Only through component-wise whiplash mechanisms like those discussed in Sec.~\ref{whiplash} would these larger-body accelerations be inherently high-risk. Repeating the above numerical for a target of mass $M=5kg$, reveals other -- potentially surprising -- and complementary insights.

\begin{figure}[H]
\includegraphics[width=0.5\textwidth]{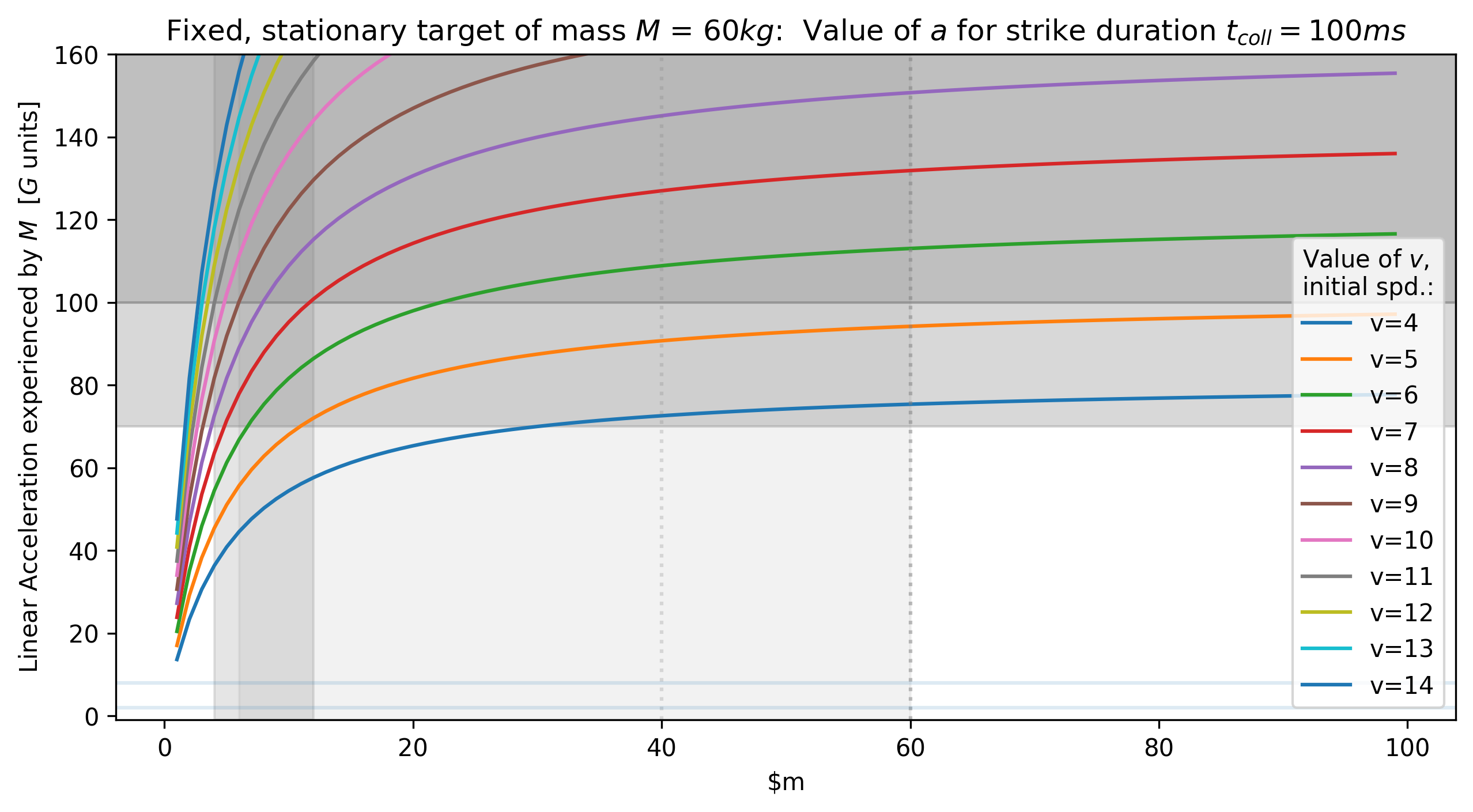}
\caption{\label{fig:accel_5kgtarget_mhoriz} Nearly any value of the incoming, effective mass $m$ above the baseline value of $5kg$ has a high chance of reaching concussion-threshold accelerations if a small target mass ($M=5kg$) is struck in such a way that the ensuing elastic collision begins and resolves in matter of $10^{-3}$ seconds or less.}
\end{figure}

The results summarized within Fig.~\ref{fig:accel_5kgtarget_mhoriz}, the results are, quintessentially, opposite to those presented in Fig.~\ref{fig:accel_60kgtarget_mhoriz}. Even incoming masses as small as the baseline of $m=5kg$ can lead unambiguously to concussion-level accelerations, with the likelihood of modeled brain trauma increasing at and above much more moderate velocities ($v\geq7\frac{m}{s}$) than those suggested by modeling collisions with the $60kg$ target, as before. While the risk of neurological trauma does increase with the value of $m$, it does so slowly, and tapers quickly after $m\sim20kg$ for all velocities $v$.

Perhaps most interestingly, despite a clear and strong $v$-dependence, with regard to determining the exact value of $a$, there is an even more profound takeaway. Namely, once strike velocity magnitudes exceed $v=5\frac{m}{s}$, the entire region of values $m\in\left[5,12.5\right]~kg$ remain accessible to 
our $70kg$ fighter, or $c\in[1,2.5)$, is able to embody strikes that pass the threshold for likely-concussive blows. The converse finding, that all larger values $m\geq25$ -- even if theoretically-inaccessible to a $70kg$ fighter -- can induce concussions for $v$ as slow as $4\frac{m}{s}$ for strikes lasting at least $10ms$ (see Wayne State Tolerance curve~\cite{gurdjian1966tolerance,tiernan2020concussion,rodriguez2023impact} for a strike of longer or shorter duration) reinforces the idea that UFC wins ``by knockout" increase linearly with weight class~\cite{ufcpi2021volume2}. Nevertheless, concussion avoidance, especially in the context of ``bare-knuckled" or minimally-gloved strikes, seems to be driven by $v$. Our results may also imply that rapid accelerations of small targets (e.g, sparring mitts) do not imply a correspondingly high $m$, and are much better-employed by observant trainers, in practice, as an inferential proxy for $v$.

\section{Summary of Key Results}
\label{summary}

Although we espouse an elastic model of martial arts, strike-induced collisions here, we do not endorse such an approximation as the best available physical approach to understanding all aspects of strike-mediated impacts. We rather wish to convey that, even within an elastic regime, our limiting bounds can be readily ascertained to support the idea that certain, manipulable variables do ``matter," insofar as determining strike outcomes -- more or less than might be expected from mathematical expressions alone -- when we take into account which sub-regions of their joint, dynamical space of possible values are, appreciably occupied during combat. We summarize:

\begin{enumerate}

  \item The effective, striking mass exerts an effect entirely \emph{independent} of that exerted by the incoming strike velocity on the post-collision velocity gained by the target mass, for a stationary target of any size.
  
  \item Recoil, or any negative change to the initial speed of a strike, is experienced by the striking implement exclusively when that strike is delivered to a \emph{target} whose mass exceeds that of the \emph{incoming} mass.
  
  \item The opposite scenario -- of no recoil, and possibly the continuation of movement of the incoming mass $m$ along the line of impact -- guarantees $m\geq M$; this may be experienced as a ``whiplash"-like effect, from the perspective of $M$, if the target is composed of coupled components far from the impact point.
  
  \item The accessible values of both \emph{absolute} changes and \emph{fold}-changes in the inducible post-collision, target velocities corresponding to any ``upstream" changes in effective striking mass tend to be of magnitudes similar to those associated with upstream changes in the initial strike speed, over a broad input space comprising nearly the full range of realizable values for $m$ and $v$ for realistic, recorded human combat.
  
  \item There exist regimes -- well within the subspaces of reasonable parameter values, representative of real human combat -- where the  kinetic energy gained by a stationary target can also be increased by the same order of magnitude as it would by varying $v$, but by increasing $m$ instead (despite that $K\propto v^2$).
  
  \item The acceleration $a$ of any stationary target $M$ \emph{small} enough to model a defender's head (i.e, the cranium and its organic contents, including the brain) -- in collision with a moving mass (striking implement) at least as massive -- can grow sufficiently large that it crosses the previously-established thresholds for inducing mild traumatic brain injury (concussion) even when strike speeds are \emph{below} the median value across experimentally-measured repetitions -- provided that the contact time is of order $\sim10^{-3}$ sec. At strike durations this short, even moderate values of $m$ are implicated in risking neurological damage.
  
  \item Any multiplicative mass transformations $m\rightarrow cm$ studied herein for which $c\in\left[1,2.5\right)$ -- and possibly even $c\in\left[1,5\right)$ -- are hypothesized to be achievable by means of training (i.e., not augmentation of the striker's \emph{overall} body mass, but modifications to a \emph{technique}). Velocity improvements are limited to a maximum scale factor of $\approx 1.67$, or certainly $b \leq 2$.
    
\end{enumerate}

\bibliography{main}

\providecommand{\noopsort}[1]{}\providecommand{\singleletter}[1]{#1}%
\begin{thebibliography}{63}
\providecommand{\natexlab}[1]{#1}
\providecommand{\url}[1]{\texttt{#1}}
\expandafter\ifx\csname urlstyle\endcsname\relax
  \providecommand{\doi}[1]{doi: #1}\else
  \providecommand{\doi}{doi: \begingroup \urlstyle{rm}\Url}\fi

\bibitem[Bluestein(2014)]{bluestein2014research}
S.J. Bluestein.
\newblock \emph{Research of Martial Arts}.
\newblock CreateSpace Independent Publishing Platform, 2014.
\newblock ISBN 9781499122510.
\newblock URL \url{https://books.google.com/books?id=E1U_BAAAQBAJ}.

\bibitem[Lee(1975)]{lee1975tao}
Bruce Lee.
\newblock \emph{Tao of Jeet Kune Do}.
\newblock Ohara Publications, Burbank, CA, 1975.
\newblock ISBN 978-0897500487.

\bibitem[Quinn(1990)]{quinn1990bouncer}
P.~Quinn.
\newblock \emph{Bouncer's Guide To Barroom Brawling: Dealing With The Sucker Puncher, Streetfighter, And Ambusher}.
\newblock Paladin Press, 1990.
\newblock ISBN 9780873645867.
\newblock URL \url{https://books.google.com/books?id=W2uAOgAACAAJ}.

\bibitem[Quinn(1996)]{quinn1996real}
P.~Quinn.
\newblock \emph{Real Fighting: Adrenaline Stress Conditioning Through Scenario-Based Training}.
\newblock Paladin Press, 1996.
\newblock ISBN 9780873648936.
\newblock URL \url{https://books.google.com/books?id=gssKAAAACAAJ}.

\bibitem[{Association of Boxing Commissions and Combative Sports}(2016)]{mmajudging2016}
{Association of Boxing Commissions and Combative Sports}.
\newblock Mma judging criteria/scoring- approved august 2, 2016, 2016.
\newblock URL \url{https://www.abcboxing.com/wp-content/uploads/2024/07/unified-mma-rules-rev-july-2024.pdf}.
\newblock Accessed May 21, 2026.

\bibitem[Eisinger(2018)]{eisinger2018}
Jordan Eisinger.
\newblock Mma wages: The determinants of ufc fighter’s salaries.
\newblock Unpublished Manuscript, UNLV Undergraduate Economics Working Paper Series, 2018.

\bibitem[Gift(2019)]{gift2019performance}
Paul Gift.
\newblock Performance bonuses and effort: Evidence from fight night awards in mixed martial arts.
\newblock \emph{International Journal of Financial Studies}, 7\penalty0 (1):\penalty0 13, 2019.

\bibitem[Stern(2026)]{whitebonus2026}
Adam Stern.
\newblock White: Ufc to double performance bonuses, add new incentive as paramount era starts, 2026.
\newblock URL \url{https://www.sportsbusinessjournal.com/Articles/2026/01/24/white-ufc-to-double-performance-bonuses-add-new-incentive-as-paramount-era-starts/}.

\bibitem[{UFC Performance Institute}(2018)]{ufcpi2018}
{UFC Performance Institute}.
\newblock A cross-sectional performance analysis and projection of the {UFC} athlete.
\newblock Technical report, Ultimate Fighting Championship (UFC), Las Vegas, NV, 2018.
\newblock URL \url{http://media.ufc.tv/ufcpi/UFCPI_Book_2018.pdf}.
\newblock Accessed 2026.

\bibitem[Lenetsky et~al.(2013)Lenetsky, Harris, and Brughelli]{lenetsky2013assessment}
Seth Lenetsky, Nigel Harris, and Matt Brughelli.
\newblock Assessment and contributors of punching forces in combat sports athletes: Implications for strength and conditioning.
\newblock \emph{Strength \& Conditioning Journal}, 35\penalty0 (2):\penalty0 1--7, 2013.

\bibitem[Monfared(2021)]{monfared2021contributing}
Saman Monfared.
\newblock Contributing factors to punching power in boxing: a narrative review summarizing determinant factors of punching power in boxing and means of improving them.
\newblock \emph{Umeå Universitet}, 2021.

\bibitem[Beattie and Ruddock(2022)]{beattie2022role}
Kris Beattie and Alan~D Ruddock.
\newblock The role of strength on punch impact force in boxing.
\newblock \emph{The Journal of Strength \& Conditioning Research}, 36\penalty0 (10):\penalty0 2957--2969, 2022.

\bibitem[Pinto et~al.(2025)Pinto, Cris{\'o}stomo, Kirk, Abi{\'a}n-Vic{\'e}n, and Monteiro]{pinto2025influence}
Manuel Pinto, Jo{\~a}o Cris{\'o}stomo, Christopher Kirk, Javier Abi{\'a}n-Vic{\'e}n, and Lu{\'\i}s Monteiro.
\newblock Influence of anthropometric characteristics and muscle performance on punch impact.
\newblock \emph{Sports}, 13\penalty0 (8):\penalty0 281, 2025.

\bibitem[Khatib et~al.(2024)Khatib, Post, Hoshizaki, and Gilchrist]{khatib2024brain}
Ali Khatib, Andrew Post, Thomas Hoshizaki, and Michael~D Gilchrist.
\newblock Brain trauma characteristics for lightweight and heavyweight fighters in professional mixed martial arts.
\newblock \emph{Sports biomechanics}, 23\penalty0 (8):\penalty0 1083--1105, 2024.

\bibitem[Beranek et~al.(2020)Beranek, Stastny, Novacek, Votapek, and Formanek]{beranek2020upper}
Vaclav Beranek, Petr Stastny, Vit Novacek, Petr Votapek, and Josef Formanek.
\newblock Upper limb strikes reactive forces in mix martial art athletes during ground and pound tactics.
\newblock \emph{International Journal of Environmental Research and Public Health}, 17\penalty0 (21):\penalty0 7782, 2020.

\bibitem[Beranek et~al.(2022)Beranek, Stastny, Novacek, S{\l}omka, and Cleather]{beranek2022performance}
Vaclav Beranek, Petr Stastny, Vit Novacek, Kajetan~J S{\l}omka, and Dan Cleather.
\newblock Performance level and strike type during ground and pound determine impact characteristics and net force variability.
\newblock \emph{Sports}, 10\penalty0 (12):\penalty0 205, 2022.

\bibitem[McGill and Marshall(2012)]{mcgill2012kettlebell}
Stuart~M McGill and Leigh~W Marshall.
\newblock Kettlebell swing, snatch, and bottoms-up carry: back and hip muscle activation, motion, and low back loads.
\newblock \emph{The Journal of Strength \& Conditioning Research}, 26\penalty0 (1):\penalty0 16--27, 2012.

\bibitem[Lee and McGill(2017)]{lee2017effect}
Benjamin Lee and Stuart McGill.
\newblock The effect of core training on distal limb performance during ballistic strike manoeuvres.
\newblock \emph{Journal of Sports Sciences}, 35\penalty0 (18):\penalty0 1768--1780, 2017.

\bibitem[Lenetsky et~al.(2015)Lenetsky, Nates, Brughelli, and Harris]{lenetsky2015effective}
Seth Lenetsky, Roy~J Nates, Matt Brughelli, and Nigel~K Harris.
\newblock Is effective mass in combat sports punching above its weight?
\newblock \emph{Human movement science}, 40:\penalty0 89--97, 2015.

\bibitem[Beranek et~al.(2023)Beranek, Votapek, and Stastny]{beranek2023force}
Vaclav Beranek, Petr Votapek, and Petr Stastny.
\newblock Force and velocity of impact during upper limb strikes in combat sports: a systematic review and meta-analysis.
\newblock \emph{Sports biomechanics}, 22\penalty0 (8):\penalty0 921--939, 2023.

\bibitem[Uthoff et~al.(2023)Uthoff, Lenetsky, Reale, Falkenberg, Pratt, Amasinger, Bourgeois, Cahill, French, and Cronin]{uthoff2023review}
Aaron Uthoff, Seth Lenetsky, Reid Reale, Felix Falkenberg, Gavin Pratt, Dean Amasinger, Frank Bourgeois, Miche{\'a}l Cahill, Duncan French, and John Cronin.
\newblock A review of striking force in full-contact combat sport athletes: Effects of different types of strength and conditioning training and practical recommendations.
\newblock \emph{Strength \& Conditioning Journal}, 45\penalty0 (1):\penalty0 67--82, 2023.

\bibitem[McGill et~al.(2010)McGill, Chaimberg, Frost, and Fenwick]{mcgill2010evidence}
Stuart~M McGill, Jon~D Chaimberg, David~M Frost, and Chad~MJ Fenwick.
\newblock Evidence of a double peak in muscle activation to enhance strike speed and force: an example with elite mixed martial arts fighters.
\newblock \emph{The Journal of Strength \& Conditioning Research}, 24\penalty0 (2):\penalty0 348--357, 2010.

\bibitem[Ishac and Eager(2021)]{ishac2021evaluating}
Karlos Ishac and David Eager.
\newblock Evaluating martial arts punching kinematics using a vision and inertial sensing system.
\newblock \emph{Sensors}, 21\penalty0 (6):\penalty0 1948, 2021.

\bibitem[Natale(2026{\natexlab{a}})]{natale2026a}
Joseph~L. Natale.
\newblock Ambulatory measurement of strike forces in modern martial arts: Recommendations and roles for novel methodologies.
\newblock Internal White Paper, 2026{\natexlab{a}}.

\bibitem[Natale(2026{\natexlab{b}})]{natale2026_shukjpotentialM}
Joseph~L. Natale.
\newblock Investigation of an exception to the rules on the upper limit of effective mass: a proposed mechanism for a new kind of martial arts strike.
\newblock In preparation., 2026{\natexlab{b}}.

\bibitem[Wolfson(2020)]{wolfson2020essential}
Richard Wolfson.
\newblock \emph{Essential University Physics (Volume 1)}, volume~1.
\newblock Pearson Education Inc., 4 edition, 2020.

\bibitem[Natale(2026{\natexlab{c}})]{natale2force}
Joseph~L. Natale.
\newblock Making sense of peak forces and contact times in amateur boxing practitioners: a new method to measure ``effective mass".
\newblock In preparation., 2026{\natexlab{c}}.

\bibitem[Kimm and Thiel(2015)]{kimm2015hand}
Dennis Kimm and David~V Thiel.
\newblock Hand speed measurements in boxing.
\newblock \emph{Procedia Engineering}, 112:\penalty0 502--506, 2015.

\bibitem[Corcoran et~al.(2024)Corcoran, Climstein, Whitting, and Del~Vecchio]{corcoran2024impact}
Daniel Corcoran, Mike Climstein, John Whitting, and Luke Del~Vecchio.
\newblock Impact force and velocities for kicking strikes in combat sports: A literature review.
\newblock \emph{Sports}, 12\penalty0 (3):\penalty0 74, 2024.

\bibitem[{Association of Boxing Commissions and Combative Sports}(2024)]{unifiedrules2024}
{Association of Boxing Commissions and Combative Sports}.
\newblock Unified rules of mixed martial arts, 2024.
\newblock URL \url{https://www.abcboxing.com/wp-content/uploads/2024/07/unified-mma-rules-rev-july-2024.pdf}.
\newblock Accessed May 21, 2026.

\bibitem[Jeremy~Nichols(2024)]{nichols2024project}
Joshua~Chin Jeremy~Nichols, Joaquin~Wang.
\newblock Exploring patterns and predictors in ufc fight outcomes, 2024.
\newblock URL \url{https://www.stat.cmu.edu/capstoneresearch/fall2024/315files_f24/team11.html}.

\bibitem[Janela(2026)]{fightmatrix2026records}
Mike Janela.
\newblock Mma fight outcomes by weight class, 2026.
\newblock URL \url{https://www.fightmatrix.com/mma-records-stats/fight-outcomes-by-weight-class/}.

\bibitem[Plagenhoef et~al.(1983)Plagenhoef, Evans, and Abdelnour]{plagenhoef1983anatomical}
Stanley Plagenhoef, F~Gaynor Evans, and Thomas Abdelnour.
\newblock Anatomical data for analyzing human motion.
\newblock \emph{Research quarterly for exercise and sport}, 54\penalty0 (2):\penalty0 169--178, 1983.

\bibitem[Zatsiorsky(1983)]{zatsiorsky1983mass}
Vladimir Zatsiorsky.
\newblock The mass and inertia characteristics of the main segments of the human body.
\newblock \emph{Biomechanics}, pages 1152--1159, 1983.

\bibitem[Kacprzak et~al.(2025)Kacprzak, Mosler, Tsos, and Wasik]{kacprzak2025biomechanics}
Jakub Kacprzak, Dariusz Mosler, Anatolij Tsos, and Jacek Wasik.
\newblock Biomechanics of punching—the impact of effective mass and force transfer on strike performance.
\newblock \emph{Applied Sciences}, 15\penalty0 (7):\penalty0 4008, 2025.

\bibitem[Walilko et~al.(2005)Walilko, Viano, and Bir]{walilko2005biomechanics}
Timothy~J Walilko, David~C Viano, and Cynthia~A Bir.
\newblock Biomechanics of the head for olympic boxer punches to the face.
\newblock \emph{British journal of sports medicine}, 39\penalty0 (10):\penalty0 710--719, 2005.

\bibitem[Stanley et~al.(2018)Stanley, Thomson, Smith, and Lamb]{stanley2018analysis}
Edward Stanley, Edward Thomson, Grace Smith, and Kevin~L Lamb.
\newblock An analysis of the three-dimensional kinetics and kinematics of maximal effort punches among amateur boxers.
\newblock \emph{International Journal of Performance Analysis in Sport}, 18\penalty0 (5):\penalty0 835--854, 2018.

\bibitem[Janela(2013)]{guiness2013}
Mike Janela.
\newblock Fan choice record august 9, 2013.
\newblock URL \url{https://www.guinnessworldrecords.com/news/2013/8/fan-choice-record-august-9-50361}.

\bibitem[Neto et~al.(2007)Neto, Magini, and Saba]{neto2007role}
Osmar~Pinto Neto, Marcio Magini, and Marcelo~MF Saba.
\newblock The role of effective mass and hand speed in the performance of kung fu athletes compared with nonpractitioners.
\newblock \emph{Journal of applied biomechanics}, 23\penalty0 (2):\penalty0 139--148, 2007.

\bibitem[Chappell et~al.(2018)Chappell, Simper, and Barker]{chappell2018nutritional}
AJ~Chappell, Trevor Simper, and ME~Barker.
\newblock Nutritional strategies of high level natural bodybuilders during competition preparation.
\newblock \emph{Journal of the International Society of Sports Nutrition}, 15\penalty0 (1):\penalty0 4, 2018.

\bibitem[Towns et~al.(2026)Towns, Cecchi, Hickey, T~O'Brien, Roberts, Pritchard, Urban, Stitzel, Grant, Zeineh, et~al.]{towns2026linear}
Jessica~A Towns, Nicholas~J Cecchi, James~W Hickey, William T~O'Brien, Spencer~SH Roberts, N~Stewart Pritchard, Jillian~E Urban, Joel~D Stitzel, Gerald~A Grant, Michael~M Zeineh, et~al.
\newblock Linear acceleration is a primary risk factor for concussion and a target for prevention.
\newblock \emph{ArXiv}, pages arXiv--2507, 2026.

\bibitem[Natale(2026{\natexlab{d}})]{natale2026_wp2}
Joseph~L. Natale.
\newblock On striking real objects and opponents: the sharpess-penetration continuum.
\newblock In preparation., 2026{\natexlab{d}}.

\bibitem[Liu and Hu(2015)]{liu2015compressive}
Yanping Liu and Hong Hu.
\newblock Compressive mechanics of warp-knitted spacer fabrics. part ii: a dynamic model.
\newblock \emph{Textile Research Journal}, 85\penalty0 (19):\penalty0 2020--2029, 2015.

\bibitem[Wasik(2009)]{wasik2009chosen}
Jacek Wasik.
\newblock Chosen aspects of physics in martial arts.
\newblock \emph{Archives of Budo}, 5:\penalty0 11--14, 02 2009.

\bibitem[Del~Vecchio et~al.(2022)Del~Vecchio, Whitting, Hollier, Keene, and Climstein]{del2022reliability}
Luke Del~Vecchio, John Whitting, Jennifer Hollier, Annabelle Keene, and Mike Climstein.
\newblock Reliability and practical use of a commercial device for measuring punch and kick impact kinetics.
\newblock \emph{Sports}, 10\penalty0 (12):\penalty0 206, 2022.

\bibitem[Ling et~al.(2016)Ling, Sanny, and Moebs]{ling2016university}
Samuel~J. Ling, Jeff Sanny, and William Moebs.
\newblock \emph{University Physics: Volume 1}.
\newblock OpenStax, Houston, Texas, 2016.
\newblock ISBN 978-1-938168-16-1.
\newblock URL \url{https://openstax.org/details/books/university-physics-volume-1}.

\bibitem[Note1()]{Note1}
Note1.
\newblock That is, nothing explicitly \protect \emph {physical}; a few \protect \emph {physiological} constraints may play a role in preventing some athletes from attaining the high-end speeds once their weights transcend a certain level, but even this pattern is not universal; moreover, it is recommended to pursue increases in $m$ through technique modifications, rather than, e.g., augmenting the size of the striking limb hypertrophically.

\bibitem[Natale(2026{\natexlab{e}})]{natale202610kfollowthrough}
Joseph~L. Natale.
\newblock What is ``follow-through"? candidate definitions for striking sports.
\newblock Unpublished Manuscript, 2026{\natexlab{e}}.

\bibitem[Jay Catherine~Kim(2026)]{kidspaperCUNY}
Joseph L.~Natale Jay Catherine~Kim.
\newblock Going beyond elastic collision models: A didactic experiment to explore collisions times and peak forces.
\newblock In preparation., 2026.

\bibitem[Mosler et~al.(2024)Mosler, Kacprzak, and Wasik]{mosler2024higher}
Dariusz Mosler, Jakub Kacprzak, and Jacek Wasik.
\newblock Higher values of force and acceleration in rear cross than lead jab: differences in technique execution by boxers.
\newblock \emph{Applied Sciences}, 14\penalty0 (7):\penalty0 2830, 2024.

\bibitem[Adamec et~al.(2021)Adamec, Hofer, Pittner, Monticelli, Graw, and Sch{\"o}pfer]{adamec2021biomechanical}
Jiri Adamec, Peter Hofer, Stefan Pittner, Fabio Monticelli, Matthias Graw, and Jutta Sch{\"o}pfer.
\newblock Biomechanical assessment of various punching techniques.
\newblock \emph{International journal of legal medicine}, 135\penalty0 (3):\penalty0 853--859, 2021.

\bibitem[Hill(1938)]{hill1938heat}
Archibald~Vivian Hill.
\newblock The heat of shortening and the dynamic constants of muscle.
\newblock \emph{Proceedings of the Royal Society of London. Series B-Biological Sciences}, 126\penalty0 (843):\penalty0 136--195, 1938.

\bibitem[Natale(2026{\natexlab{f}})]{natale2026b}
Joseph~L. Natale.
\newblock The 10,000 fists study: Insights from the first 2k fists.
\newblock In preparation., 2026{\natexlab{f}}.

\bibitem[Broglio et~al.(2010)Broglio, Schnebel, Sosnoff, Shin, Feng, He, and Zimmerman]{broglio2010biomechanical}
Steven~P Broglio, Brock Schnebel, Jacob~J Sosnoff, Sunghoon Shin, Xingdong Feng, Xuming He, and Jerrad Zimmerman.
\newblock The biomechanical properties of concussions in high school football.
\newblock \emph{Medicine and science in sports and exercise}, 42\penalty0 (11):\penalty0 2064, 2010.

\bibitem[Tiernan et~al.(2020)Tiernan, Meagher, O’Sullivan, O’Keeffe, Kelly, Wallace, Doherty, Campbell, Liu, and Domel]{tiernan2020concussion}
Stephen Tiernan, Aidan Meagher, David O’Sullivan, Eoin O’Keeffe, Eoin Kelly, Eugene Wallace, Colin~P Doherty, Matthew Campbell, Yuzhe Liu, and August~G Domel.
\newblock Concussion and the severity of head impacts in mixed martial arts.
\newblock \emph{Proceedings of the Institution of Mechanical Engineers, Part H: Journal of engineering in medicine}, 234\penalty0 (12):\penalty0 1472--1483, 2020.

\bibitem[Lee and McGill(2014)]{lee2014striking}
Benjamin Lee and Stuart McGill.
\newblock Striking dynamics and kinetic properties of boxing and mma gloves.
\newblock \emph{Revista de Artes Marciales Asi{\'a}ticas}, 9\penalty0 (2), 2014.

\bibitem[Broglio et~al.(2012{\natexlab{a}})Broglio, Eckner, and Kutcher]{broglio2012field}
Steven~P Broglio, James~T Eckner, and Jeffery~S Kutcher.
\newblock Field-based measures of head impacts in high school football athletes.
\newblock \emph{Current opinion in pediatrics}, 24\penalty0 (6):\penalty0 702--708, 2012{\natexlab{a}}.

\bibitem[Broglio et~al.(2012{\natexlab{b}})Broglio, Surma, and Ashton-Miller]{broglio2012high}
Steven~P Broglio, Tyler Surma, and James~A Ashton-Miller.
\newblock High school and collegiate football athlete concussions: a biomechanical review.
\newblock \emph{Annals of biomedical engineering}, 40\penalty0 (1):\penalty0 37--46, 2012{\natexlab{b}}.

\bibitem[Hons and Tiernan(2020)]{hons2020measurement}
A~Hons and Mr~Stephen Tiernan.
\newblock Measurement and simulation of head impacts in mixed martial arts.
\newblock \emph{Measurement}, 2020:\penalty0 08--11, 2020.

\bibitem[Tierney(2024)]{tierney2024concussion}
Gregory Tierney.
\newblock Concussion biomechanics, head acceleration exposure and brain injury criteria in sport: a review.
\newblock \emph{Sports biomechanics}, 23\penalty0 (11):\penalty0 1888--1916, 2024.

\bibitem[Gurdjian et~al.(1966)Gurdjian, Roberts, and Thomas]{gurdjian1966tolerance}
Elisha~S Gurdjian, VL~Roberts, and L~Murray Thomas.
\newblock Tolerance curves of acceleration and intracranial pressure and protective index in experimental head injury.
\newblock \emph{Journal of Trauma and Acute Care Surgery}, 6\penalty0 (5):\penalty0 600--604, 1966.

\bibitem[Rodriguez-Millan et~al.(2023)Rodriguez-Millan, Rubio, Burpo, Olmedo, Loya, Parker, and Migu{\'e}lez]{rodriguez2023impact}
Marcos Rodriguez-Millan, I~Rubio, FJ~Burpo, A~Olmedo, JA~Loya, KK~Parker, and MH~Migu{\'e}lez.
\newblock Impact response of advance combat helmet pad systems.
\newblock \emph{International Journal of Impact Engineering}, 181:\penalty0 104757, 2023.

\bibitem[{UFC Performance Institute}(2021)]{ufcpi2021volume2}
{UFC Performance Institute}.
\newblock A cross-sectional performance analysis and projection of the {UFC} athlete, volume 2.
\newblock Technical report, Ultimate Fighting Championship, Las Vegas, NV, 2021.
\newblock URL \url{https://www.ufc.com/news/ufc-performance-institute-publishes-pivotal-follow-groundbreaking-mma-study-read}.

\end{thebibliography}
\bibliographystyle{unsrtnat}

\end{document}